\documentclass[journal]{IEEEtran}
\usepackage{multirow}
\usepackage{graphicx}
\usepackage{amsmath}
\usepackage[psamsfonts]{amssymb}
\usepackage{amsxtra}
\usepackage{threeparttable}
\usepackage{subfigure}
\usepackage{caption}
\usepackage{epstopdf}
\usepackage{stfloats}
\usepackage{algorithm}
\usepackage{algorithmicx}
\usepackage{algpseudocode}

\ifCLASSINFOpdf
\else
\fi
\begin{document}
%
\title{Fake Colorized Image Detection}
%
%
%

\author{Yuanfang Guo,~\IEEEmembership{Member,~IEEE,}
        Xiaochun Cao,~\IEEEmembership{Senior Member,~IEEE.}
        Wei Zhang,~\IEEEmembership{Member,~IEEE,}
        Rui Wang,~\IEEEmembership{Member,~IEEE,}
\thanks{Yuanfang Guo, Xiaochun Cao, Wei Zhang and Rui Wang are with the State Key Laboratory of Information Security, Institute of Information Engineering, Chinese Academy of Sciences, Beijing 100093, China (email: guoyuanfang@iie.ac.cn/eeandyguo@connect.ust.hk, caoxiaochun@iie.ac.cn, wzhang@iie.ac.cn, wangrui@iie.ac.cn).}
\thanks{Xiaochun Cao is also with the School of Cyber Security, University of Chinese Academy of Sciences, Beijing 100049, China.}}

\maketitle

\begin{abstract}
Image forensics aims to detect the manipulation of digital images. Currently, splicing detection, copy-move detection and image retouching detection are drawing much attentions from researchers. However, image editing techniques develop with time goes by. One emerging image editing technique is colorization, which can colorize grayscale images with realistic colors. Unfortunately, this technique may also be intentionally applied to certain images to confound object recognition algorithms. To the best of our knowledge, no forensic technique has yet been invented to identify whether an image is colorized. We observed that, compared to natural images, colorized images, which are generated by three state-of-the-art methods, possess statistical differences for the hue and saturation channels. Besides, we also observe statistical inconsistencies in the dark and bright channels, because the colorization process will inevitably affect the dark and bright channel values. Based on our observations, i.e., potential traces in the hue, saturation, dark and bright channels, we propose two simple yet effective detection methods for fake colorized images: Histogram based Fake Colorized Image Detection (FCID-HIST) and Feature Encoding based Fake Colorized Image Detection (FCID-FE). Experimental results demonstrate that both proposed methods exhibit a decent performance against multiple state-of-the-art colorization approaches.
\end{abstract}


%
\IEEEpeerreviewmaketitle

\section{Introduction}\label{sec:intro}

The rapid proliferation of image editing technologies has increased both the ease with which images can be manipulated and the difficulty in distinguishing between altered and natural images. In addition to the conventional image editing techniques such as splicing [\ref{farid2009}], copy-move [\ref{Li2015}] and retouching [\ref{cao2014}], more image editing techniques, such as colorization [\ref{Larsson2016}] and image generation [\ref{goodfellow2014gan}], are proposed. Since these types of image editing techniques generate new content with/without references, we denote them as the generative image editing techniques.

Although image editing techniques can provide significant aesthetic or entertainment value, they may also be used with malicious intent. In general, the various image editing approaches employ different mechanisms. Splicing and copy-move techniques usually manipulate part of the image and perform object-level changes. Image retouching techniques usually change the images via a variety of mechanisms. For example, contrast enhancement adjusts the contrast of the image, while image inpainting usually fills the holes in images according to the image content. Among the generative image editing techniques, image generation usually produces a meaningful image from a noise vector with/without some additional information such as text or a class label. Colorization, on the other hand, usually colorizes images with visually plausible colors, which may cause misjudgment when specific objects/scenes must be identified/tracked.

Fortunately, numerous image forensic technologies have been developed in the past decades. According to their mechanisms and applications, they can be categorized into two classes, active techniques and passive techniques. The active techniques usually refer to watermarking techniques [\ref{huang2016rdh}-\ref{wang2017multiw}], which embed authentication information to the to-be-protected images. When the integrities of these images demand verification, watermark extraction procedures are performed and the extracted watermarks are compared to the original watermark to detect forgeries. Since the active techniques require the watermark to be embedded prior to detection, the applications, in practice, are limited.

In contrast, passive image forgery detection approaches, to which our proposed methods belong, usually detect the manipulations to the input images directly. Traditionally, passive image forgery (editing) detection techniques have mainly focused on splicing detection [\ref{farid2009}], copy-move detection [\ref{Li2015}] and image retouching detection [\ref{cao2014}]. To the best of our knowledge, no method has yet been developed to detect the fake images generated by generative image editing techniques. If the images are examined by humans, the cost increases drastically as the number of to-be-examined images increases. Obviously, detection via human eyes is incompetent for the big data era. On the other hand, the conventional image forgery detection techniques are designed with different assumptions that may not be appropriate for generative fake image detection. Therefore, generative fake image detection demands specific studies and designs.

\begin{figure*}
  \centering
  \subfigure[]{
    \includegraphics[width=170mm]{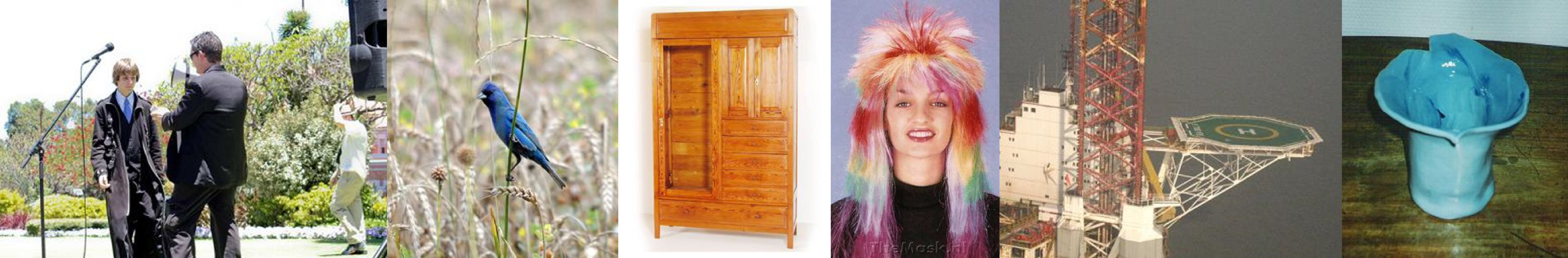}\label{fig:realexamples}}
  \subfigure[]{
    \includegraphics[width=170mm]{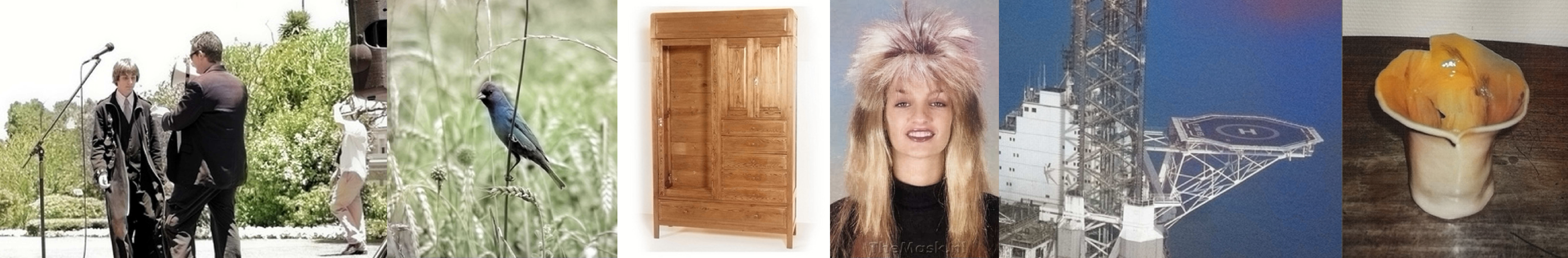}\label{fig:fakeexamples}}
\caption{(a) Real images. (b) Fake colorized images. \label{fig:motivation}}
\end{figure*}

Among the different generative image forgery techniques, colorization already achieves excellent performances. As Fig. \ref{fig:motivation} shows, fake colorized images, which are generated by a state-of-the-art algorithm [\ref{Larsson2016}], are visually indistinguishable, if no ground-truth images exist for comparison. Therefore, the necessity to develop a scheme for fake colorized image detection increases quickly. In this paper, we aim to address this new problem by providing feasible solutions. Specifically, we propose two simple yet effective schemes for detecting fake colorized images, which are generated by fully automatic colorization methods. The contributions are summarized as follows:

\begin{description}
  \item[1:] ~ We observe that fake colorized images and their corresponding natural images exhibit statistical differences, which can be further utilized as detection traces, in both color channels and image prior. The color channels involved are the hue and saturation channels, while the exploited extreme channels prior is proposed in our recent work [\ref{yang2017ecp}].
  \item[2:] ~ According to the statistical differences in the color channels and image priors, we propose a fake colorized image detection scheme, named Histogram based Fake Colorized Image Detection (FCID-HIST), by proposing four detection features. Each feature calculates the most distinctive bin and the total variation of the normalized histogram distribution for hue, saturation, dark and bright channels, respectively.
  \item[3:] ~ To better utilize the statistical information of the training images, we consider exploiting the divergences inside different moments of the data vectors and propose a fake colorized image detection scheme, named Feature Encoding based Fake Colorized Image Detection (FCID-FE), by modeling the created four-dimensional samples with a Gaussian Mixture Model (GMM)[\ref{far2005}] and encoding the samples into Fisher feature vectors [\ref{perronnin2007fv}].
  \item[4:] ~ In the experiments, the two proposed methods demonstrate a decent performance in various tests for detecting fake images generated by three state-of-the-art colorization methods.
\end{description}

The rest of the paper is organized as follows. Section \ref{sec:background} presents the necessary background. Section \ref{sec:method} introduces the proposed work. Section \ref{sec:results} describes the experimental results in various tests and analyzes the proposed methods. Finally, Section \ref{sec:conclusion} summarizes the paper and discusses future work.

\section{Background}\label{sec:background}

In this section, the conventional forgery detection techniques and the colorization techniques are reviewed accordingly.

\subsection{Review of Forgery Detection}

Forgery detection [\ref{qureshi2015survey}] has been studied for decades. In general, forgery detection explores different characteristics of images and attempts to find traces to analyze. As mentioned above, most of the traditional forgery detection techniques can be categorized into three classes, copy-move detection, splicing detection and image retouching detection.

Copy-move detection relies on finding duplicated regions in a tampered image. Intuitively, these techniques tend to seek an appropriate feature in a certain domain, such that the detection can be performed via searching the most similar two units (such as patches). Different methods usually exploit different features. [\ref{fridrich2003}] explores features in the frequency domain by dividing the image into overlapping blocks and detects the copy-move forgery via matching the quantized DCT coefficients. [\ref{Yu2010}] performs a rotation invariant detection based on the Fourier-Mellin Transform. [\ref{kirchner2013}] localizes the duplicated regions based on the Zernike moments, which exhibit the rotation invariance property, of small image blocks. [\ref{kirchner2013}] reports decent results especially when the duplicated regions are smooth. [\ref{Amerini2011}] employs the famous SIFT feature [\ref{lowe2004sift}] to detect multiple duplicated regions and estimates the geometric transformation performed by the copy-move operation. [\ref{Liu2014}] presents a SIFT based detection method by matching the SIFT features via a broad first search neighbors clustering algorithm and further distinguishing the duplicated origins from the tampered regions via CFA features. [\ref{Jiantao2016}] introduces a hierarchical SIFT-based keypoint matching technique to solve a drawback of previous keypoint matching based detection techniques, which tends to give poor performances when the copy-moved regions are small or smooth. Although copy-move detection technologies have been developed rapidly, they cannot be directly applied to the fake colorized image detection because no copy-move operations exist in the fake colorized images.

Splicing detection usually detects the manipulated regions which originate from different source images. Different from copy-move detection, these approaches detect the tampered regions with various traces (features), which usually reveal the inconsistencies between the tampered regions and the unchanged regions. Currently, splicing detection can be classified into four categories, compression-based methods, camera-based methods, physics-based methods and geometry-based methods, according to their mechanisms.

Compression-based methods assume that the spliced region and the original image have undergone different types of image compression and may exhibit different compression artifacts. For example, [\ref{bianchi2012}] considers the DCT coefficient distributions of each $8\times8$ block and computes the tampering probability. By considering the advantages and disadvantages of different block sizes, [\ref{korus2016}] constructs a multiscale scheme, employs the Benford's law at each level and fuses the results together to obtain a final localization map. Unfortunately, the compression-based methods are not appropriate for fake colorized image detection because the assumption may not always hold.

Camera-based methods consider traces left on the image during the capturing process. [\ref{ferrara2012}] detects the existences of the CFA artifacts, which are due to the demosaicking process in the CFA cameras, and thus obtains the localization map. [\ref{chierchia2014}] exploits the photo-response non-uniformity noises (i.e., the sensor noises) of the camera to distinguish the tampered regions from the original ones. [\ref{korus2017}] also considers the photo-response non-uniformity noises and a multiscale framework to conduct a multiscale analysis and detects small forgeries more accurately. Even if the camera-based methods can be employed to detect the fake colorized images, their robustness is incompetent because the sensor noises and the artifacts can easily be affected by noises and some common post-processing operations such as compression.

Physics-based methods perform detection based on different physics phenomenon inconsistencies. [\ref{bahrami2015}] considers the blur type inconsistency between the spliced region and the original image to localize the tampered region. [\ref{carvalho2016}] explores the illuminant-based transform spaces and combines different image descriptors such as color, shape and texture to detect forged regions. Since the fake colorized images to be examined in this paper are forged for the whole image, these inconsistencies cannot be utilized to detect the fake colorized images.

Geometry-based methods utilize the geometry information inside images for detection. [\ref{zhang2009}] explores detecting the compositions with the two-view geometrical constraints. [\ref{zhang2010}] considers the planar homographies in the test images and adopts graph-cut algorithm to obtain the final localization map. Unfortunately, since the geometrical characteristics are hardly manipulated in the fake colorized images, the geometry-based methods will also fail to detect the colorized images.

Image retouching detection usually considers that the original images are restored or enhanced. For example, [\ref{trung2014}] is designed for detecting the inpainted images by considering the similarities, distances and number of identical pixels among different blocks. [\ref{cao2014}] calculates the histograms and detects via the peak/gap artifacts induced from contrast enhancement. These techniques can hardly be applied to the new problem because their mechanisms are specially designed for their own assumptions.

Table \ref{tab:sumexistingdetection} shows a summary of existing forensic techniques. Although many detection technologies have been developed, they are currently not directly applicable to the detection of images manipulated by generative methods. Specially designed techniques are necessary to address the detection of fake colorized images.

\begin{table*}
\begin{center}
\caption{Summary of the existing fake image detection approaches} \label{tab:sumexistingdetection}
\begin{tabular}{|c|c|c|c|}
  \hline
    Category & Method & Core mechanism & Potential result of detecting colorized images \\
  \hline
    \multirow{6}*{Copy-move detection} & [\ref{fridrich2003}] & Quantized DCT coefficients & Not applicable \\
                                       & [\ref{Yu2010}] & Fourier-Mellin Transform & Not applicable \\
                                       & [\ref{kirchner2013}] & Zernike moments & Not applicable \\
                                       & [\ref{Amerini2011}] & SIFT feature & Not applicable \\
                                       & [\ref{Liu2014}] & SIFT \& CFA features & Not applicable \\
                                       & [\ref{Jiantao2016}] & Hierarchical SIFT-based keypoint matching & Not applicable \\
  \hline
    \multirow{9}*{Splicing detection} & [\ref{bianchi2012}] & DCT coefficient distributions of each block & Not applicable \\
                                      & [\ref{korus2016}] & Multiscale scheme based on Benford's law & Not applicable \\
                                      & [\ref{ferrara2012}] & CFA artifacts & Possible but with low robustness \\
                                      & [\ref{chierchia2014}] & PRNU noises & Possible but with low robustness \\
                                      & [\ref{korus2017}] & Multiscale scheme based on PRNU noises & Possible but with low robustness \\
                                      & [\ref{bahrami2015}] & Blur type inconsistency & Not applicable \\
                                      & [\ref{carvalho2016}] & Illuminant-based transform spaces & Not applicable \\
                                      & [\ref{zhang2009}] & Two-view geometrical constraints & Not applicable \\
                                      & [\ref{zhang2010}] & Planar homographies & Not applicable \\

  \hline
    \multirow{2}*{Image retouching detection} & [\ref{trung2014}] & Block similarities and distances & Not applicable \\
                                              & [\ref{cao2014}] & Peak/gap artifacts & Not applicable \\
  \hline
\end{tabular}
\end{center}
\end{table*}

\subsection{Review of Colorization}

Colorization, a term describing the color adding process to grayscale images, was firstly introduced by Wilson Markle in 1970. However, this area began to develop rapidly in the 21st century. Colorization techniques can be categorized into the following types: scribble-based, example-based and fully automatic.

Scribble-based methods are supervised techniques in which users begin assigning colors to pixels in the grayscale image. The milestone work [\ref{Levin2004}], which assumes that the neighboring pixels with similar intensities should have similar colors, is proposed at first. Various other approaches have been proposed in succession, such as [\ref{Pang2013}], which constructs dictionaries for color and textures via sparse representation and colorizes the images accordingly.

Example-based methods[\ref{Charpiat2008}][\ref{chen2016}] usually require the users to supervise the system by providing reference color image(s) similar to the greyscale image. The system then transfers the colors in the reference color image(s) to the target greyscale image by searching for similar patterns/objects. The performances of these methods are dependent on the quality of the reference image(s). If the divergence between the greyscale image and the reference image(s) is high, the colorized result may be unsatisfactory.

In contrast with the supervised approaches above, fully automatic methods require no supervision when performing the colorization task. [\ref{Cheng2015}] trains a neural network and predicts the chrominance values by considering the pixel patch, DAISY and semantic features. [\ref{lizuka2016tog}] colorizes the images by jointly utilizing the local and global priors with an end-to-end network. [\ref{Larsson2016}] proposes a state-of-the-art approach, which exploits the hypercolumn to utilize both low-level and semantic representations, and colorizes the images in the Hue-Chroma-Lightness (HCL) color space. [\ref{zhang2016eccv}] calculates the statistical distributions of the chrominance information in the LAB space and introduces a classification-style colorization approach based on a deep network.

These techniques are briefly summarized in Table \ref{tab:sumcolorization}. Due to the high performances of the fully automatic colorization techniques, we focus on the detection of the fake colorized images generated via these techniques in this paper.

\begin{table*}
\begin{center}
\caption{Summary of the existing colorization approaches} \label{tab:sumcolorization}
\begin{tabular}{|c|c|c|c|}
  \hline
    Category & Method & Core mechanism & Side-information \\
  \hline
    \multirow{2}*{Scribble-based method} & [\ref{Levin2004}] & Neighboring pixels with similar intensities should have similar colors & User scribbles \\
                                         & [\ref{Pang2013}] & Construct color and texture dictionaries & User scribbles \\
  \hline
    \multirow{2}*{Example-based method} & [\ref{Charpiat2008}] & Global optimization of colors at pixel-level & Reference color image \\
                                        & [\ref{chen2016}] & Propagating the learned dictionaries & Reference color image \\
  \hline
    \multirow{4}*{Fully automatic method} & [\ref{Cheng2015}] & Network with pixel patch, DAISY and semantic features & None \\
                                          & [\ref{lizuka2016tog}] & End-to-End network with local and global priors & None \\
                                          & [\ref{Larsson2016}] & End-to-End network with Hypercolumn & None \\
                                          & [\ref{zhang2016eccv}] & Classification-style colorization in LAB space& None \\
  \hline
\end{tabular}
\end{center}
\end{table*}

\section{Methodology}\label{sec:method}

The rapid progress in colorization technologies has enabled colorized images to be visually indistinguishable from natural images. State-of-the-art colorization methods are already capable of misleading human observers in the subjective tests [\ref{zhang2016eccv}]. To distinguish the fake colorized images from the natural images, we study the statistics of the fake colorized images, which are generated by three state-of-the-art methods [\ref{Larsson2016}][\ref{zhang2016eccv}][\ref{lizuka2016tog}], and propose two simple yet effective detection schemes, FCID-HIST and FCID-FE.

\subsection{Observations and Statistics}\label{sec:sd}

\begin{figure*}
  \centering
  \subfigure[]{
    \includegraphics[width=5.5cm]{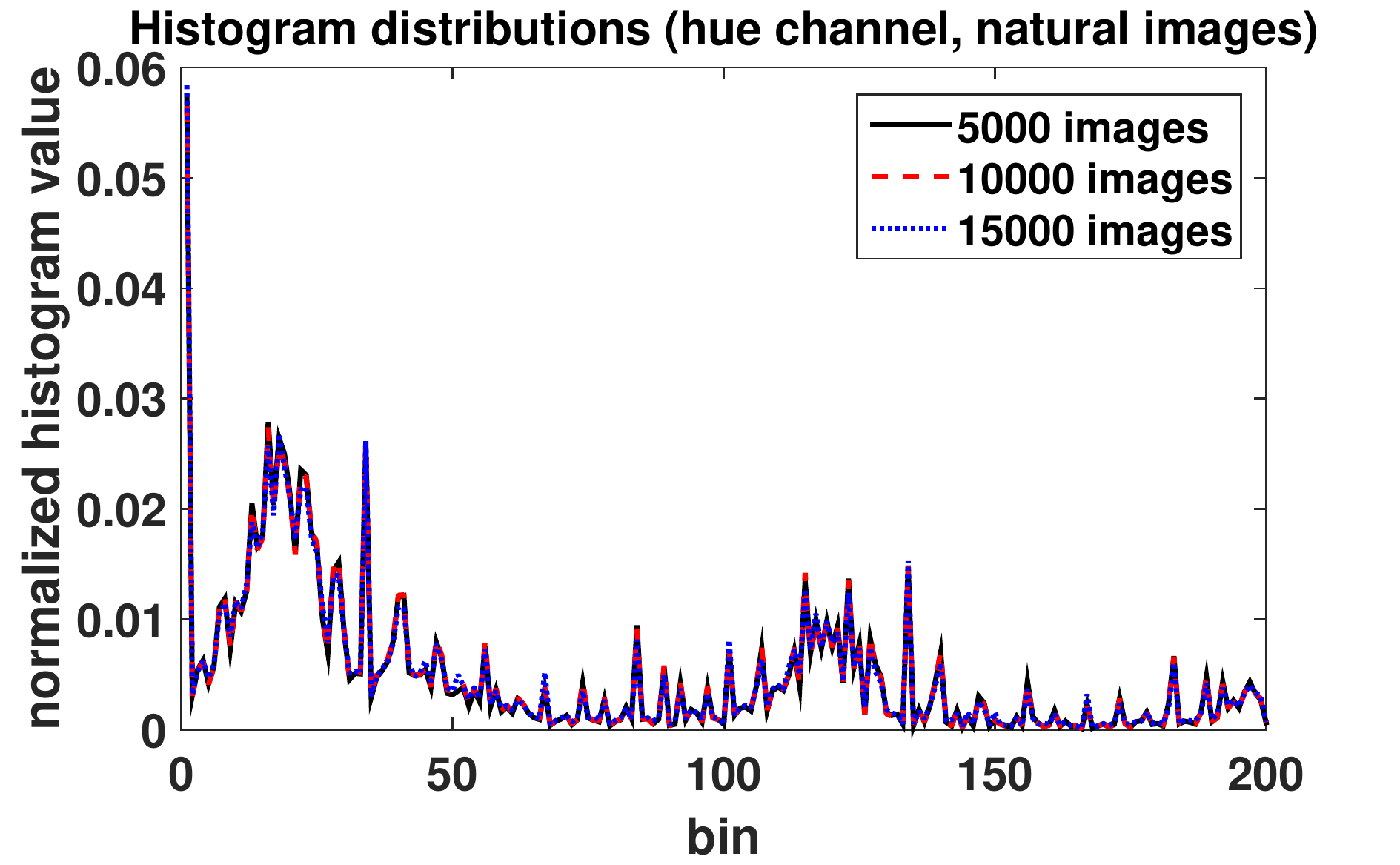}\label{fig:truehuehist}}
  \subfigure[]{
    \includegraphics[width=5.5cm]{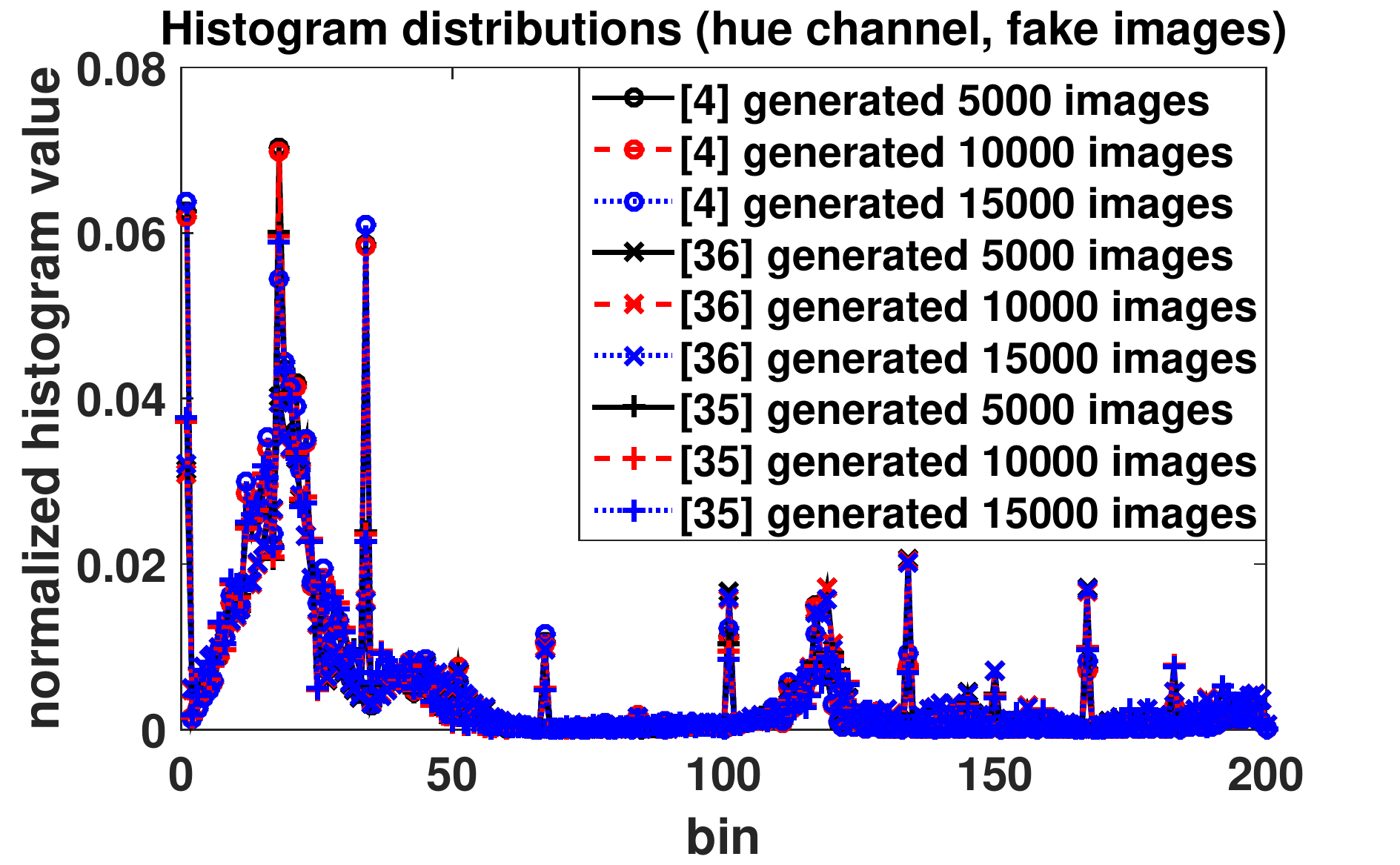}\label{fig:fakehuehist}}
  \subfigure[]{
    \includegraphics[width=5.5cm]{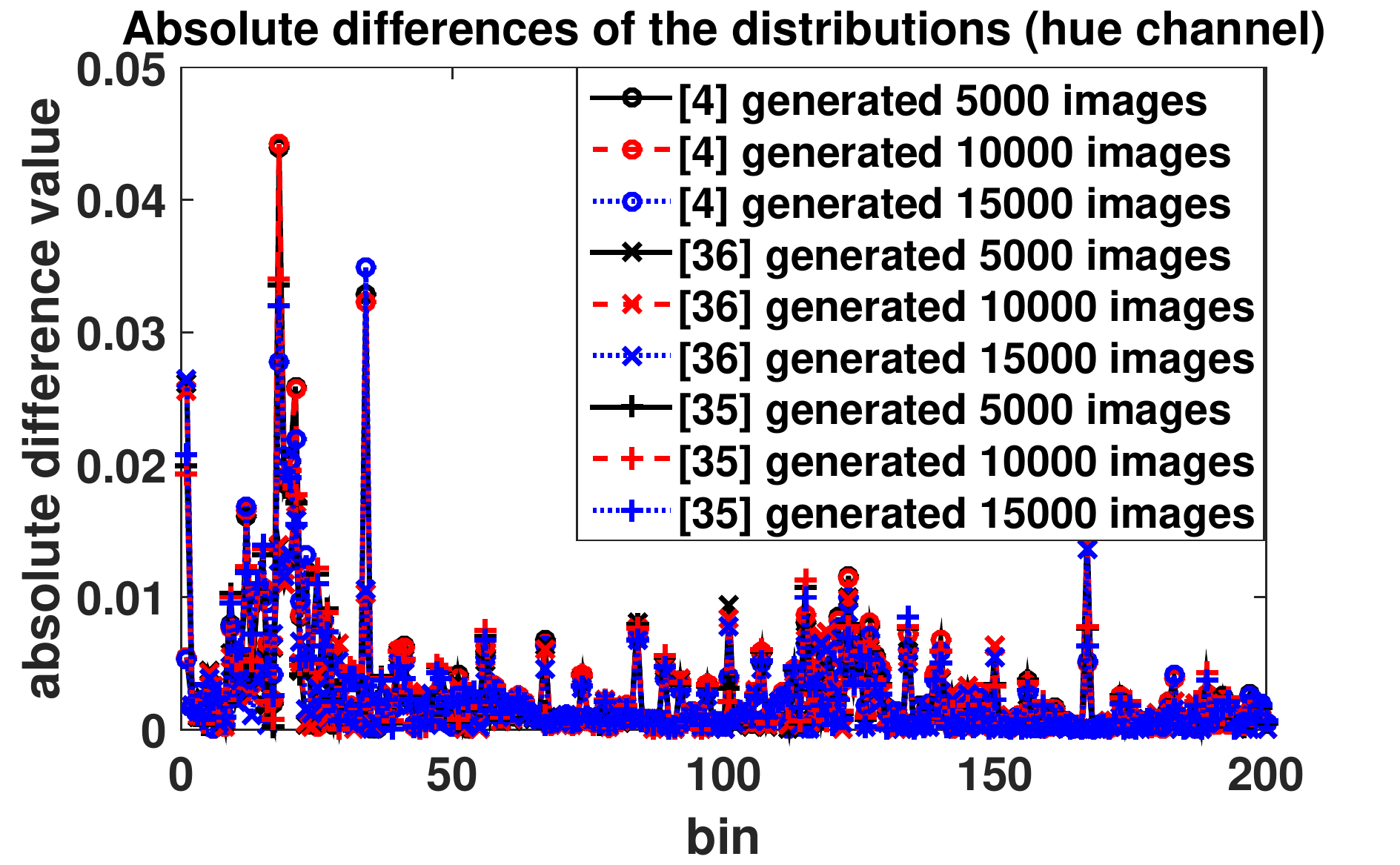}\label{fig:diffhuehist}}\\
  \subfigure[]{
    \includegraphics[width=5.5cm]{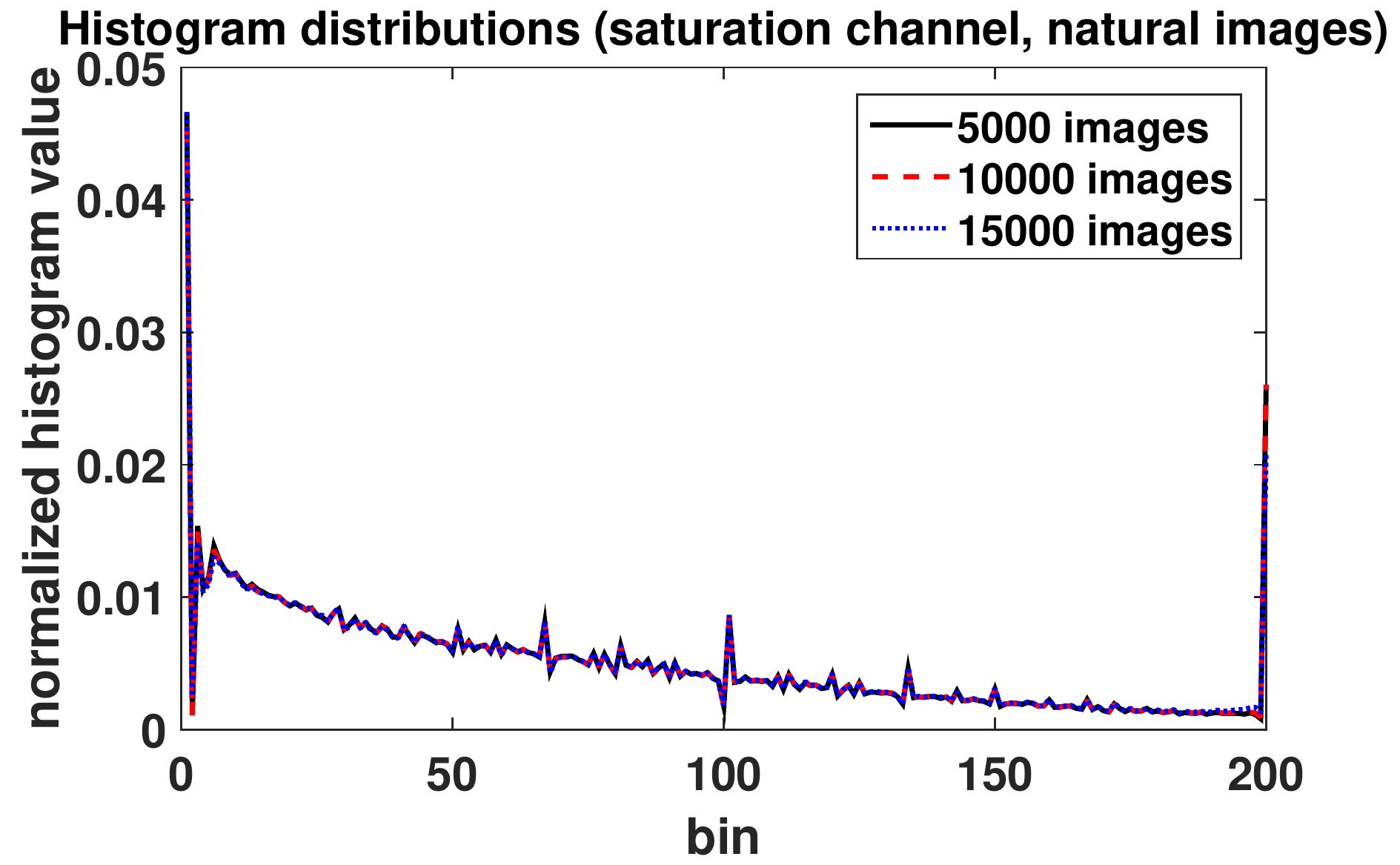}\label{fig:truesathist}}
  \subfigure[]{
    \includegraphics[width=5.5cm]{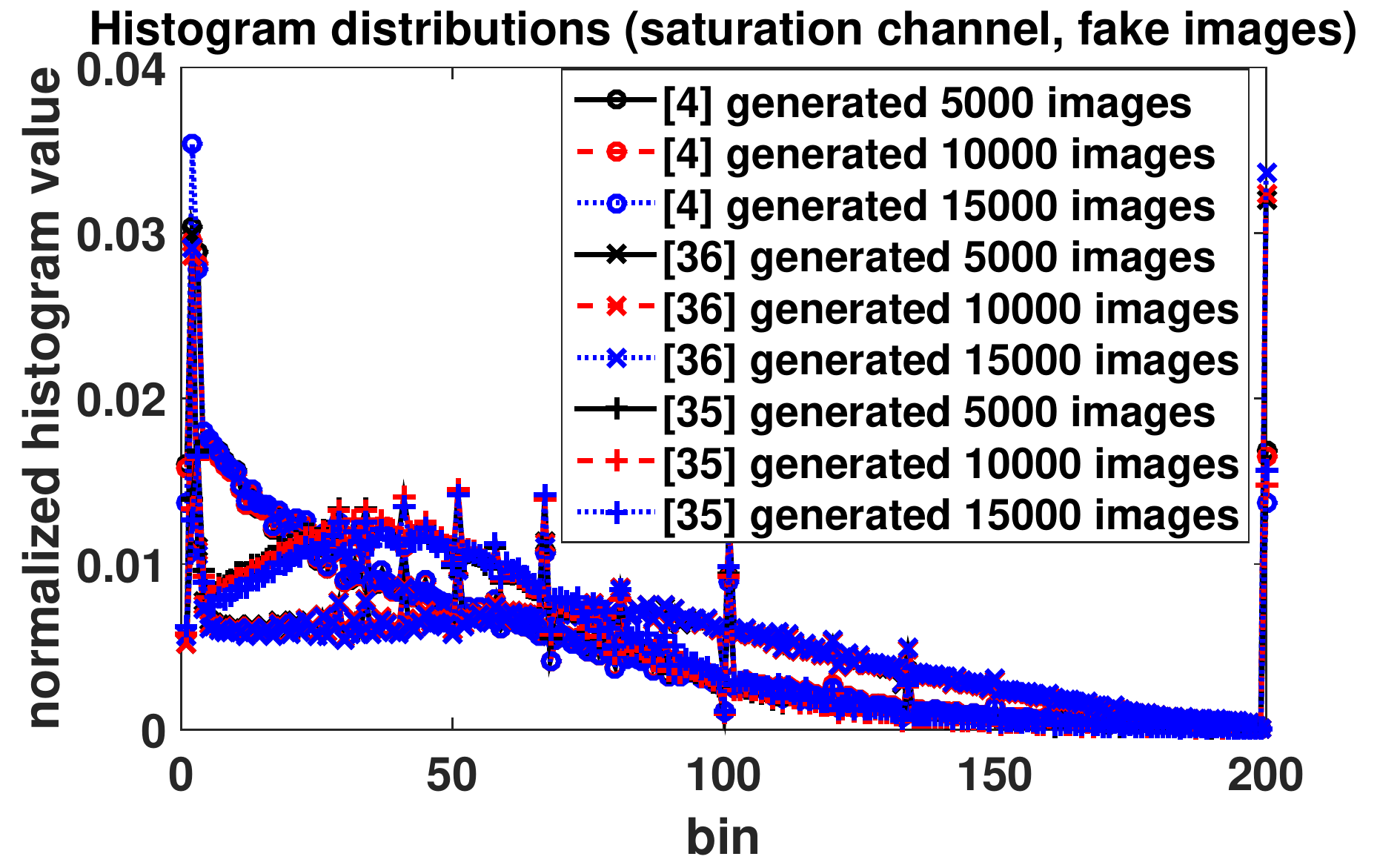}\label{fig:fakesathist}}
  \subfigure[]{
    \includegraphics[width=5.5cm]{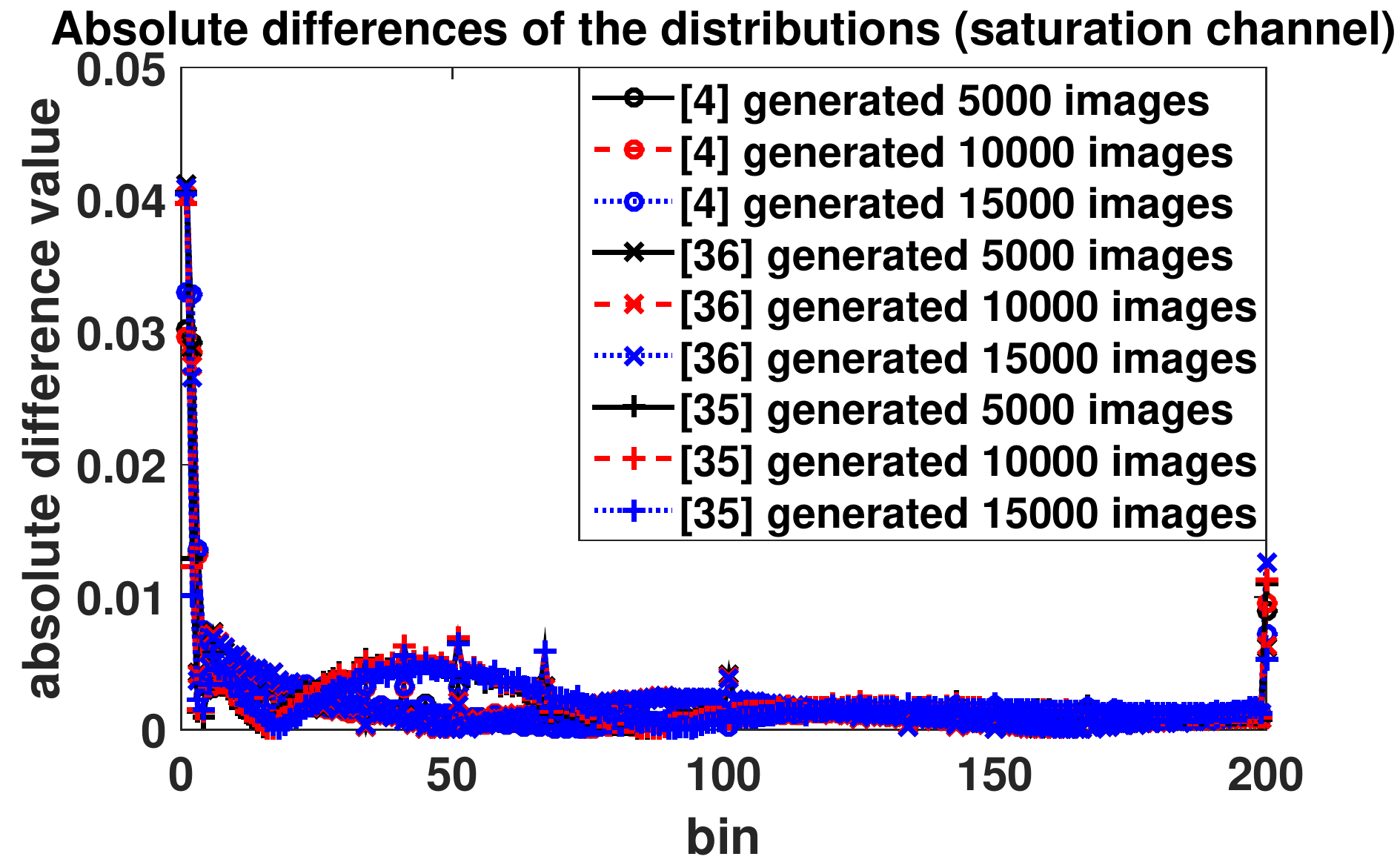}\label{fig:diffsathist}}
\caption{(a) Normalized histogram distribution of the hue channel (natural images). (b) Normalized histogram distribution of the hue channel (fake images). (c) Absolute differences of the distributions in (a) and (b). (d) Normalized histogram distribution of the saturation channel (natural images). (e) Normalized histogram distribution of the saturation channel (fake images). (f) Absolute differences of the distributions in (d) and (e). \label{fig:HShistogram}}
\end{figure*}

According to our observation, the colorized images tend to possess less saturated colors, and the colorization method favors some colors over others, though these differences are difficult to visually detect. Since the Hue-Saturation-Value (HSV) color space separately represents the chrominance information in the hue and saturation channel, we calculate the normalized histograms (each containing 200 bins) of the hue and saturation channel in 15000 natural images and their corresponding fake colorized images, separately, as shown in Fig. \ref{fig:HShistogram}.

As shown in Fig. \ref{fig:HShistogram}, the statistics of the natural and fake colorized images are different in both the hue and saturation channels, and there also exist statistical differences (especially for the peaks in the histograms) among the fake images generated by different colorization methods. For the hue channel, the histogram of the fake images tends to be more smooth and possesses more significant peaks compared to the natural images. For the saturation channel, the histogram of the fake images also exhibits different peak values and variances compared to the histogram of the natural images. These statistics indicate that the fake images favor different colors and possess saturation differences compared to the natural images. Therefore, the natural and fake colorized images are statistically identifiable, though the fake colorized images seemed visually indistinguishable.

In addition to the statistical differences in the color channels, differences also exist in some image priors because they are not considered explicitly in the colorization process even though the deep neural networks possesses good generalization ability. In this paper, we exploit our recently proposed extreme channels prior (ECP) [\ref{yang2017ecp}], which consists of the dark channel prior (DCP) [\ref{he2011dcp}] and the bright channel prior (BCP). Intuitively, DCP assumes that the dark channel of a natural image is close to zero, while BCP assumes that the bright channel of a natural image is close to $255$. The dark channel $I_{dc}$ and bright channel $I_{bc}$ of an image $I$ are defined as shown by Eq. \ref{eq:dcp} and \ref{eq:bcp}, respectively.
\begin{align}\label{eq:dcp}
    I_{dc}(x)=\underset{y\in\Omega(x)}{\text{min}}\big(\underset{c_p\in{(r,g,b)}}{\text{min}}I_{c_p}(y)\big),
\end{align}
\begin{equation} \label{eq:bcp}
    I_{bc}(x)=\underset{y\in\Omega(x)}{\text{max}}\big(\underset{c_p\in{(r,g,b)}}{\text{max}}I_{c_p}(y)\big),
\end{equation}
where $x$ stands for the pixel location, $I_{c_p}$ denotes a color channel of $I$ and $\Omega(x)$ represents the local patch centered at location $x$. Note that the local patch sizes here are identical to the settings in [\ref{yang2017ecp}].

\begin{figure*}
  \centering
  \subfigure[]{
    \includegraphics[width=5.5cm]{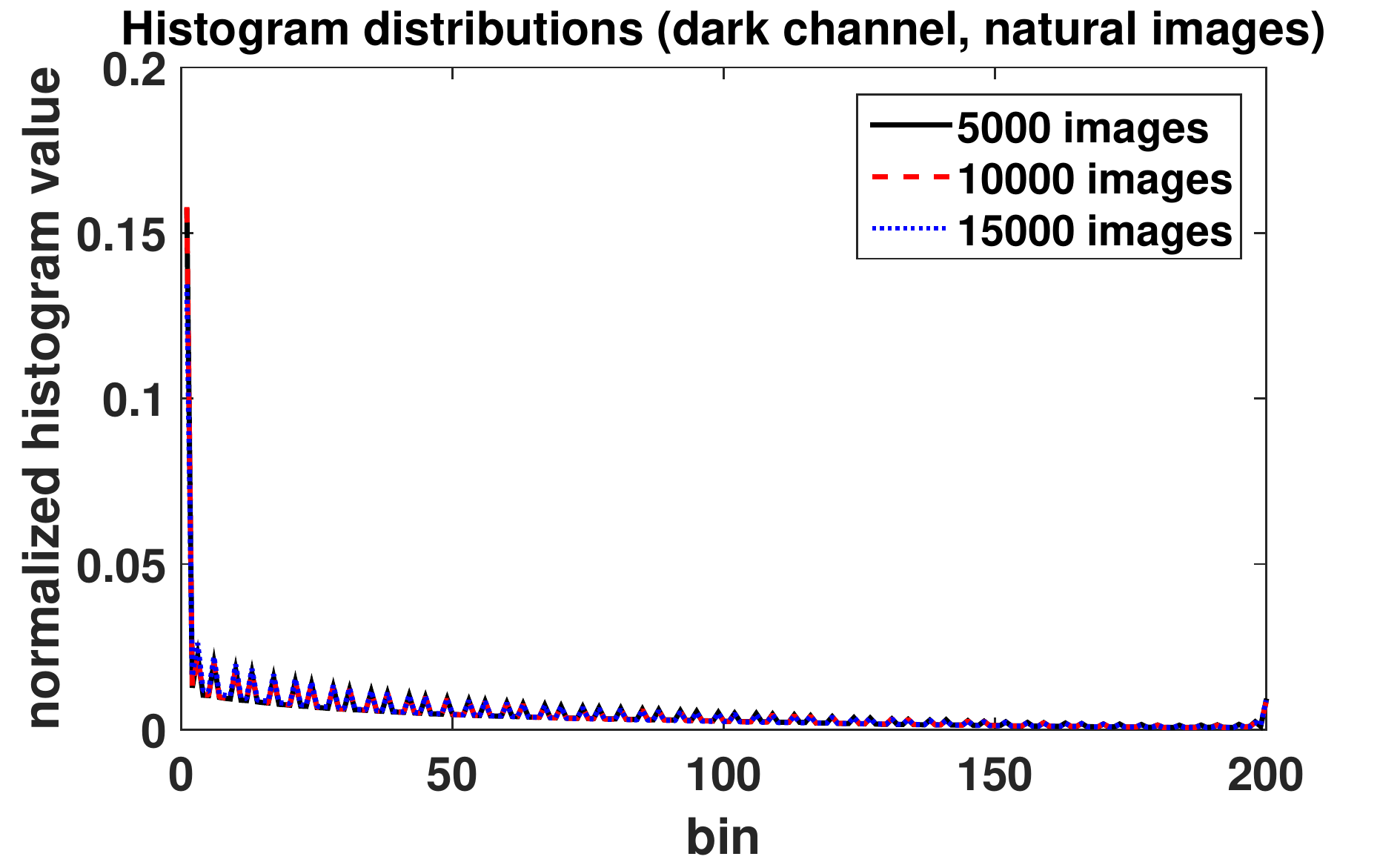}\label{fig:truedcphist}}
  \subfigure[]{
    \includegraphics[width=5.5cm]{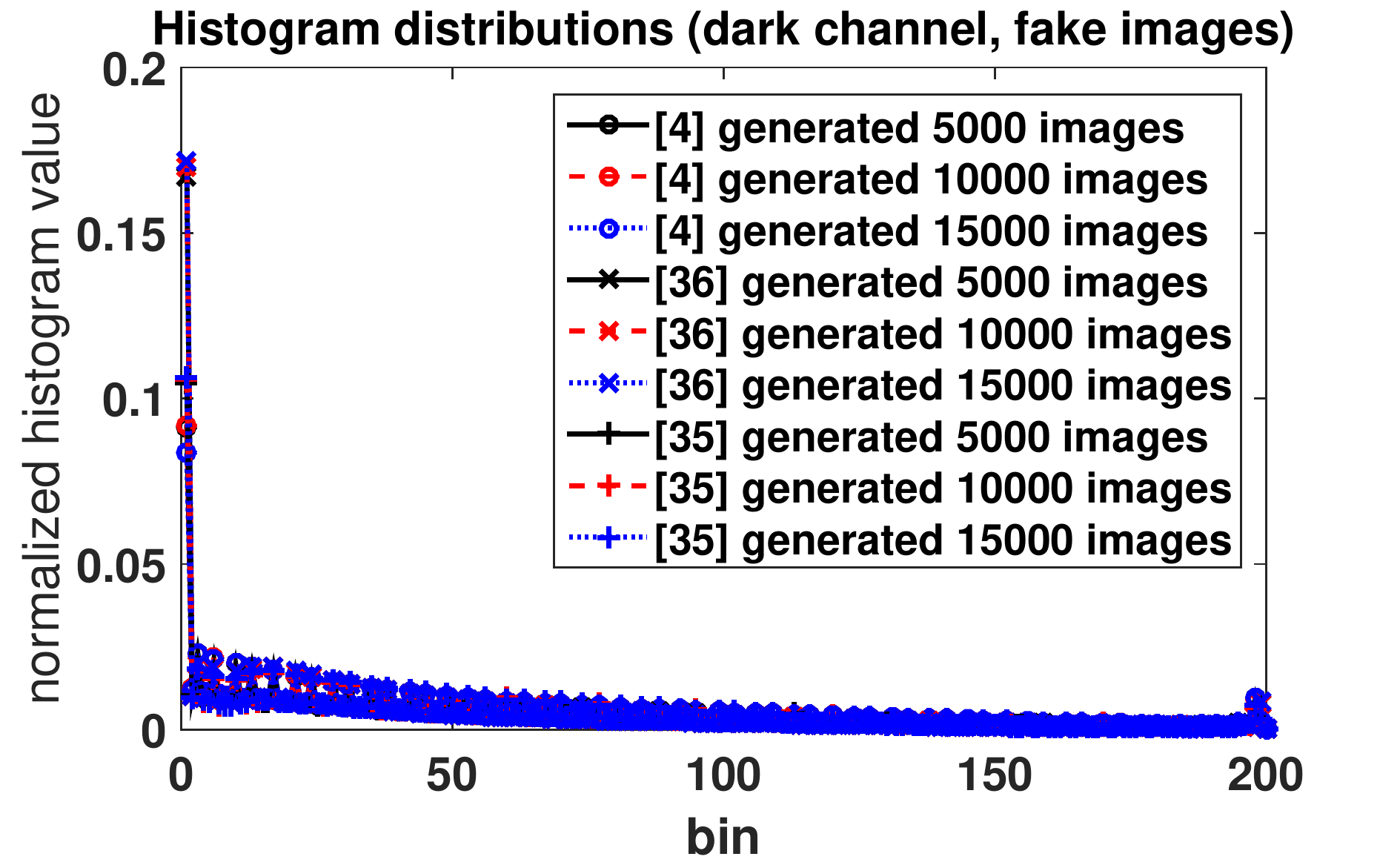}\label{fig:fakedcphist}}
  \subfigure[]{
    \includegraphics[width=5.5cm]{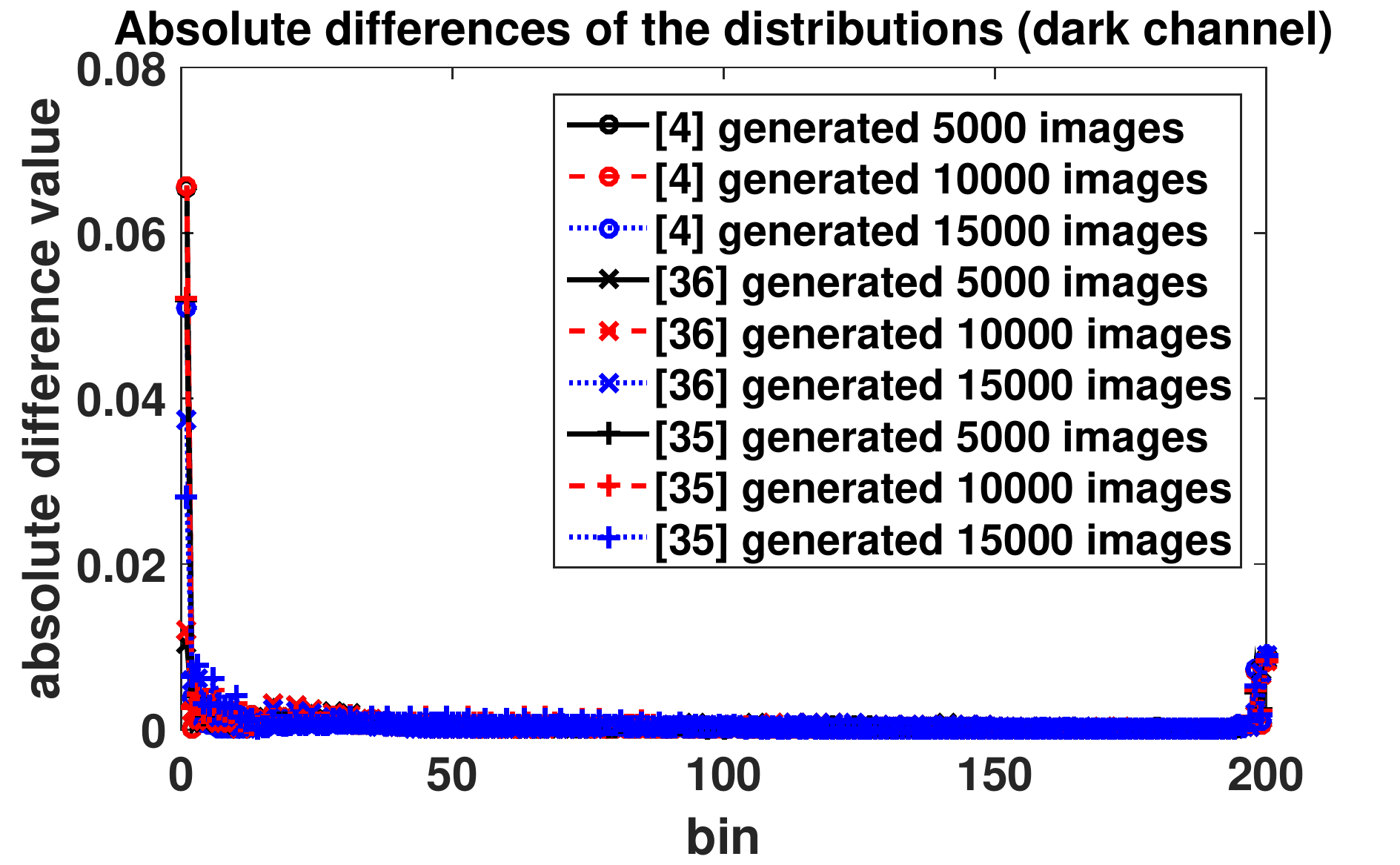}\label{fig:diffdcphist}}\\
  \subfigure[]{
    \includegraphics[width=5.5cm]{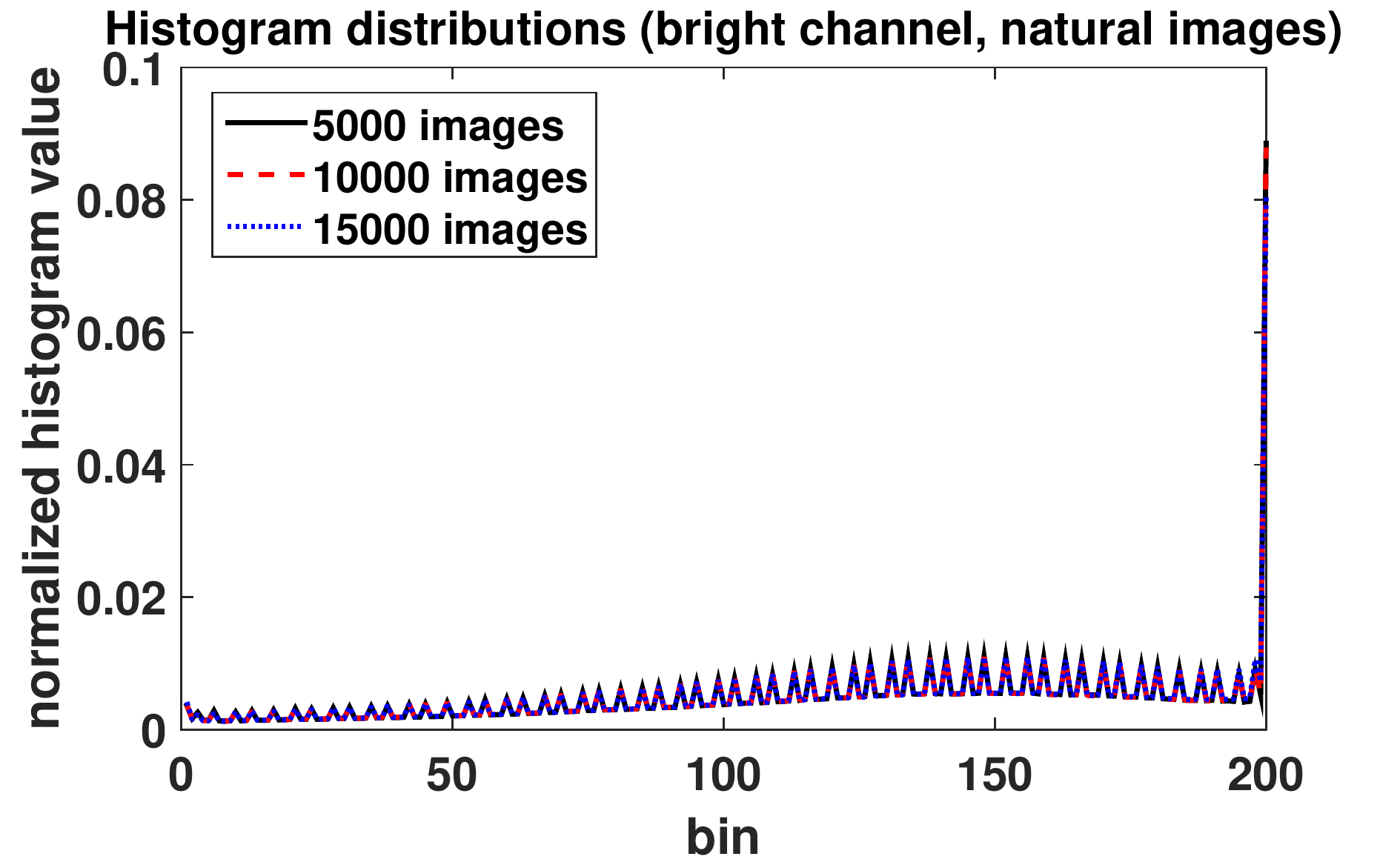}\label{fig:truebcphist}}
  \subfigure[]{
    \includegraphics[width=5.5cm]{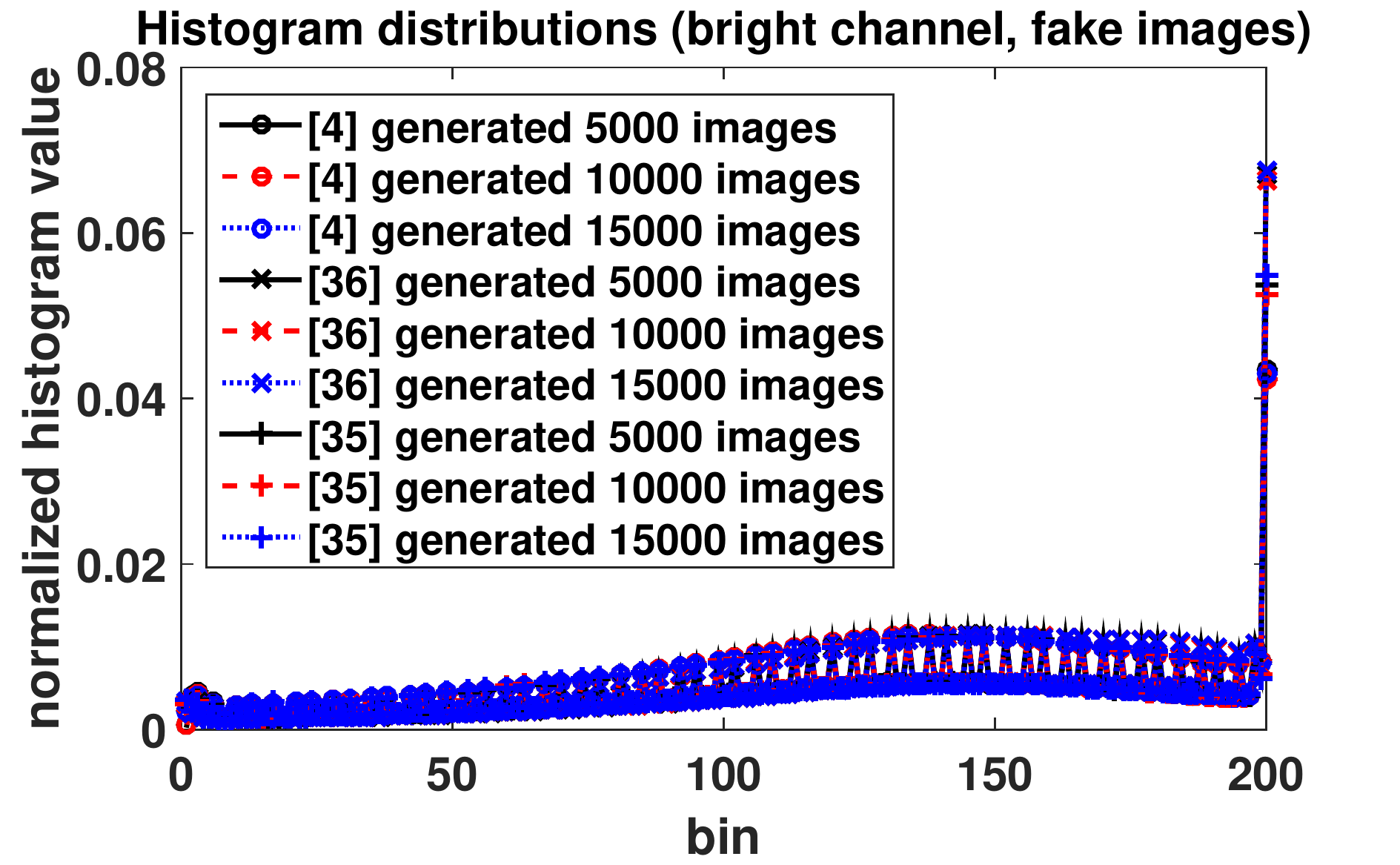}\label{fig:fakebcphist}}
  \subfigure[]{
    \includegraphics[width=5.5cm]{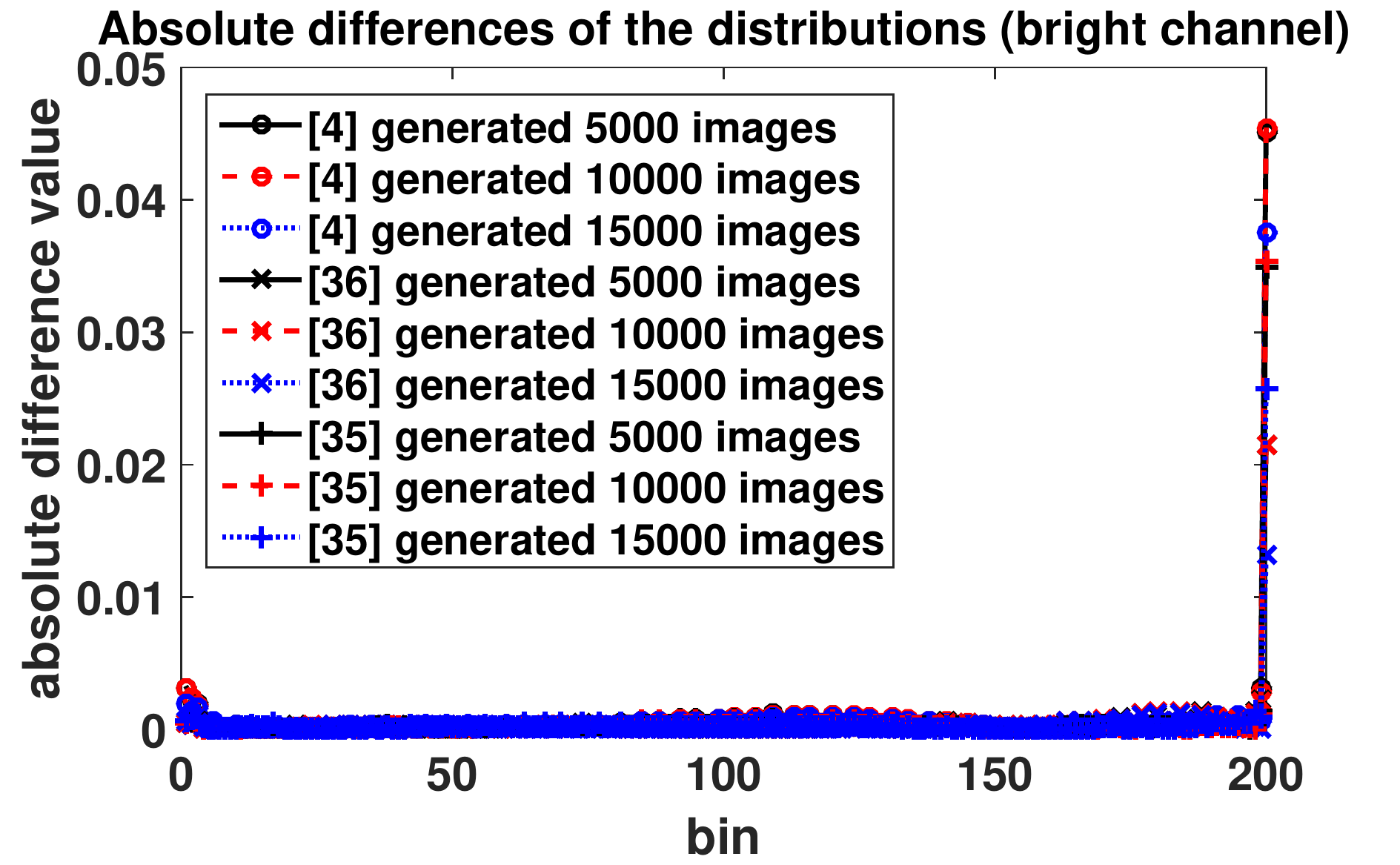}\label{fig:diffbcphist}}
\caption{(a) Normalized histogram distribution of the dark channel (natural images). (b) Normalized histogram distribution of the dark channel (fake images). (c) Absolute differences of the distributions in (a) and (b). (d) Normalized histogram distribution of the bright channel (natural images). (e) Normalized histogram distribution of the bright channel (fake images). (f) Absolute differences of the distributions in (d) and (e). \label{fig:ECPhistogram}}
\end{figure*}

By calculating the histograms of the dark channel and bright channel of 15000 natural images and their corresponding fake colorized images separately, Fig. \ref{fig:ECPhistogram} presents the expected differences, especially for the peak values, and supports our observations above.

\subsection{FCID-HIST}\label{sec:baseline}

By exploiting the existing statistical differences, we propose the Histogram based Fake Colorized Image Detection (FCID-HIST) method to detect fake colorized images.

In FCID-HIST, four detection features, the hue feature $F_h$, the saturation feature $F_s$, the dark channel feature $F_{dc}$ and the bright channel feature $F_{bc}$, are proposed to detect forgeries.

The hue feature is constructed from the normalized hue channel histogram distributions. Let $K_h$ be the total number of bins in each normalized hue channel histogram distribution. We define $Dist_{h,n}$ and $Dist_{h,f}$ as the normalized hue channel histogram distribution for the natural and fake training images, respectively, and $Dist_{h}^\alpha$ as the corresponding histogram for the $\alpha$th input image, which can be either a training or testing image.

Intuitively, to differentiate the fake colorized images from the natural images, the distinctive features should reveal the largest divergences between the two types of images. (Note that, in this paper, the Euclidean distance is employed to calculate the divergences.) Therefore, we select the most distinctive bin $Dist_{h}^\alpha(\upsilon_h)$, whose two corresponding bins in $Dist_{h,n}$ and $Dist_{h,f}$ give the largest divergence between the two histogram distributions, as part of the hue feature, as follows
\begin{align}\label{eq:huedistribution1}
    F_{h}^\alpha(1) = Dist_{h}^\alpha(\upsilon_h)
\end{align}
where the index of the most distinctive bin $\upsilon_h$ for the hue channel is calculated via Eq. \ref{eq:huedistribution2}.
\begin{align}\label{eq:huedistribution2}
    \upsilon_h & = \underset{x}{\text{argmax}}||Dist_{h,n}(x)-Dist_{h,f}(x)||_2 \nonumber\\
               & = \underset{x}{\text{argmax}}|Dist_{h,n}(x)-Dist_{h,f}(x)|
\end{align}

The distributions $Dist_{h,n}$ and $Dist_{h,f}$ also vary differently with respect to the bins. We account for this difference in the hue feature by computing the first order derivative of the normalized hue channel histogram distribution $DistD_{h}^\alpha(l)=Dist_{h}^\alpha(l+1)-Dist_{h}^\alpha(l)$ to capture the varying trend of the histogram distribution. This total variation is calculated as Eq. \ref{eq:huedistribution3} shows.
\begin{align}\label{eq:huedistribution3}
    F_{h}^\alpha(2) = \overset{K_h-1}{\underset{l=1}{\sum}}|DistD_{h}^\alpha(l)|
\end{align}

The proposed hue feature is then formed by combining Eq. \ref{eq:huedistribution1} with Eq. \ref{eq:huedistribution3} into a vector, as Eq. \ref{eq:huefeature} demonstrates.
\begin{align}\label{eq:huefeature}
    F_h^\alpha = [F_{h}^\alpha(1) \text{ } F_{h}^\alpha(2)]
\end{align}

Similarly, the saturation feature $F_{s}^\alpha$, the dark channel feature $F_{dc}^\alpha$ and the bright channel feature $F_{bc}^\alpha$ can be constructed by utilizing the normalized histogram distributions ($Dist_{s,n}$, $Dist_{s,f}$), ($Dist_{dc,n}$, $Dist_{dc,f}$), and ($Dist_{bc,n}$, $Dist_{bc,f}$) for the saturation, bright, and dark channels of the training images respectively.

In the same manner as Eq. \ref{eq:huedistribution2}, the indexes for the most distinctive bins $\upsilon_s$, $\upsilon_{dc}$ and $\upsilon_{bc}$ can be calculated by Eq. \ref{eq:otherdistribution1}.
\begin{align}\label{eq:otherdistribution1}
    \upsilon_{c_h} = \underset{x}{\text{argmax}}|Dist_{c_h,n}(x)-Dist_{c_h,f}(x)|, \text{ }c_h={s,dc,bc}
\end{align}

Then, the most distinctive bins for each feature can be calculated via Eq. \ref{eq:otherdistribution2}.
\begin{align}\label{eq:otherdistribution2}
    & F_{c_h}^\alpha(1) = Dist_{c_h}^\alpha\big(\underset{x}{\text{argmax}}|Dist_{c_h,n}(x)-Dist_{c_h,f}(x)|\big), \\
    & \text{ }c_h={s,dc,bc} \nonumber
\end{align}
where $Dist_{c_h}^\alpha$ represents the normalized $c_h$ channel histogram distribution of the $\alpha$th input image.

The total variation of each distribution is computed via Eq. \ref{eq:otherdistribution3}.
\begin{align}\label{eq:otherdistribution3}
    F_{c_h}^\alpha(2) = \overset{K_{c_h}-1}{\underset{l=1}{\sum}}|DistD_{c_h}^\alpha(l)|, \text{ }c_h={s,dc,bc}
\end{align}
where $K_{c_h}$ stands for the total number of bins in each normalized $c_h$ channel histogram distribution and $DistD_{c_h}^\alpha$ denotes the first order derivative of the normalized $c_h$ channel histogram distribution.

Then, the features are formed as shown in Eq. \ref{eq:otherfeature}.
\begin{align}\label{eq:otherfeature}
    F_{c_h}^\alpha = [F_{c_h,0}^\alpha \text{ } F_{c_h,1}^\alpha], \text{ }c_h={s,dc,bc}
\end{align}

With all the features calculated, the final detection feature $F_{HIST}^\alpha$ for the $\alpha$th input image can be constructed via Eq. \ref{eq:finalfeature}.
\begin{align}\label{eq:finalfeature}
    F_{HIST}^\alpha = [F_h^\alpha \text{ } F_s^\alpha \text{ } F_{dc}^\alpha \text{ } F_{bc}^\alpha]
\end{align}

After the detection feature is calculated, FCID-HIST employs the supporting vector machine (SVM)[\ref{chang2011svm}] for training and detecting the fake colorized images. The FCID-HIST algorithm is summarized as shown in Algorithm \ref{alg:fcidb}.
For convenience, we let $K_h=K_s=K_{dc}=K_{bc}$ in this paper.

\begin{algorithm}[t]
\caption{FCID-HIST}
\raggedright
\textbf{Training Stage:}\\
\textbf{Input:} $N_{1}$ natural and fake colorized training images, the corresponding labels $L_{r,HIST}$, $K_h$, $K_s$, $K_{dc}$, $K_{bc}$, SVM parameters\\
\textbf{Output:} $\upsilon_h$, $\upsilon_s$, $\upsilon_{dc}$, $\upsilon_{bc}$, trained SVM classifier\\
\begin{algorithmic}[1]
\State Compute $Dist_{h,n}$, $Dist_{s,n}$, $Dist_{dc,n}$, $Dist_{bc,n}$
\State Compute $Dist_{h,f}$, $Dist_{s,f}$, $Dist_{dc,f}$, $Dist_{bc,f}$
\State Compute $\upsilon_h$, $\upsilon_s$, $\upsilon_{dc}$, $\upsilon_{bc}$ \Comment refer to Eq. \ref{eq:huedistribution2} and \ref{eq:otherdistribution1}
\For{$i=1$ to $N_{1}$}
\State Compute $Dist_{h}^i$, $Dist_{s}^i$, $Dist_{dc}^i$, $Dist_{bc}^i$
\State Compute $F_{h}^i(1)$, $F_{s}^i(1)$, $F_{dc}^i(1)$, $F_{bc}^i(1)$ \Comment refer to Eq. \ref{eq:huedistribution1} and \ref{eq:otherdistribution2}
\State Compute $F_{h}^i(2)$, $F_{s}^i(2)$, $F_{dc}^i(2)$, $F_{bc}^i(2)$ \Comment refer to Eq. \ref{eq:huedistribution3} and \ref{eq:otherdistribution3}
\State Compute $F_{h}^i$, $F_{s}^i$ , $F_{dc}^i$, $F_{bc}^i$ \Comment refer to Eq. \ref{eq:huefeature} and \ref{eq:otherfeature}
\State Compute $F_{HIST}^i$ \Comment refer to Eq. \ref{eq:finalfeature}
\EndFor
\State Train SVM with $F_{HIST}$, $L_{r,HIST}$ and SVM parameters
\end{algorithmic}
\vspace{0.3cm}
\textbf{Testing Stage:}\\
\textbf{Input:} ~~\ 
$N_{2}$ test images, $K_h$, $K_s$, $K_{dc}$, $K_{bc}$, $\upsilon_h$, $\upsilon_s$, $\upsilon_{dc}$, $\upsilon_{bc}$, trained SVM classifier\\
\textbf{Output:} ~~\ 
Detection labels $L_{e,HIST}$\\
\begin{algorithmic}[1]
\For{$i=1$ to $N_{2}$}
\State Compute $Dist_{h}^i$, $Dist_{s}^i$, $Dist_{dc}^i$, $Dist_{bc}^i$
\State Compute $F_{h}^i(1)$, $F_{s}^i(1)$, $F_{dc}^i(1)$, $F_{bc}^i(1)$ \Comment refer to Eq. \ref{eq:huedistribution1} and \ref{eq:otherdistribution2}
\State Compute $F_{h}^i(2)$, $F_{s}^i(2)$, $F_{dc}^i(2)$, $F_{bc}^i(2)$ \Comment refer to Eq. \ref{eq:huedistribution3} and \ref{eq:otherdistribution3}
\State Compute $F_{h}^i$, $F_{s}^i$ , $F_{dc}^i$, $F_{bc}^i$ \Comment refer to Eq. \ref{eq:huefeature} and \ref{eq:otherfeature}
\State Compute $F_{HIST}^i$ \Comment refer to Eq. \ref{eq:finalfeature}
\State Obtain $L_{e,HIST}(i)$ with $F_{HIST}^i$ and the trained SVM clasifier
\EndFor
\end{algorithmic}
\label{alg:fcidb}
\end{algorithm}

\subsection{FCID-FE}\label{sec:advanced}

Although FCID-HIST gives a decent performance in the experiments, which are demonstrated in the latter section, these features may not fully utilize the statistical differences between the natural and fake colorized images because the distributions are modeled channel by channel. Therefore, we propose another scheme, Feature Encoding based Fake Colorized Image Detection (FCID-FE), to better exploit the statistical information by jointly modeling the data distribution and exploiting the divergences inside different moments of the distribution.

Let $I_{h}^\beta$, $I_{s}^\beta$, $I_{dc}^\beta$ and $I_{bc}^\beta$ be the hue, saturation, dark and bright channels of a training image respectively, where $\beta$ is the index of the training image. Then, we create a training sample set $\Phi$ via Eq. \ref{eq:sample}.
\begin{align}\label{eq:sample}
    & \Phi\big((z-1)*i*j+(i-1)*j+j\big) \nonumber \\
    & = [I_{h}^\beta(i,j)\text{ }I_{s}^\beta(i,j)\text{ }I_{dc}^\beta(i,j)\text{ }I_{bc}^\beta(i,j)]
\end{align}

In contrast to the histogram modeling, FCID-FE models the sample data distribution $G$ with a Gaussian Mixture Model (GMM) [\ref{far2005}] as shown in Eq. \ref{eq:gmm}.
\begin{align}\label{eq:gmm}
    G(\Phi|\Theta) = \overset{N}{\underset{n=1}{\sum}}\log{p(\Phi_n|\Theta)}
\end{align}
where $N$ is the number of samples in $\Phi$, $\Theta$ stands for the parameter set of the constructed GMM and $\Theta$ is defined as shown in Eq. \ref{eq:paramset}.
\begin{align}\label{eq:paramset}
    \Theta = {\omega_a, \mu_a, \sigma_a}, \text{ }a=1,...,N_m, \text{ }\overset{N_m}{\underset{n=1}{\sum}}\omega_a=1
\end{align}
where $\omega_a$ represents the weight, $\mu_a$ stands for the mean value vector, $\sigma_a$ denotes the covariance matrix and $N_m$ is the number of Gaussian distributions in the distribution model.

Then, the likelihood of $\Phi_n$ being modeled by the GMM $\Theta$ can be represented by Eq. \ref{eq:gmm2}.
\begin{align}\label{eq:gmm2}
    p(\Phi_n|\Theta) = \overset{N_m}{\underset{m=1}{\sum}}\log{\omega_m p_m(\Phi_m|\Theta)}
\end{align}
where $p_m(\Phi_m|\Theta)$ is defined by Eq. \ref{eq:gmm3}.
\begin{align}\label{eq:gmm3}
    p_m(\Phi_m|\Theta) = \frac{\exp{[-(1/2)(\Phi_m-\mu_a)^T\sigma_a^{-1}(\Phi_m-\mu_a)]}}{(2\pi)^{N_v/2}|\sigma_a|^{1/2}}
\end{align}
where $N_v$ denotes the number of dimensions of each sample vector. Then, GMM can be constructed by determining the parameter set $\Theta$.


With the determined GMM, FCID-FE utilizes different moments of the distribution and encodes each subset $\Phi^\beta$ of the sample vectors, which belongs to each training image, into training Fisher vectors [\ref{perronnin2007fv}] as Eq. \ref{eq:fv} shows.
\begin{align}\label{eq:fv}
    F_{FE}^\beta = [\frac{\lambda_1\delta G(\Phi^\beta|\Theta)}{\delta\omega_a}\text{ }
           \frac{\lambda_2\delta G(\Phi^\beta|\Theta)}{\delta\mu_{a,v}}\text{ }
           \frac{\lambda_3\delta G(\Phi^\beta|\Theta)}{\delta\sigma_{a,v}}]
\end{align}
where $v=1,2,...,N_v$ and $\lambda_1$, $\lambda_2$ and $\lambda_3$ are defined in Eqs. \ref{eq:par1}-\ref{eq:par3}.
\begin{align}\label{eq:par1}
    \lambda_1 = \big(N(\frac{1}{\omega_a}+\frac{1}{\omega_1})\big)^{-1/2}
\end{align}
\begin{align}\label{eq:par2}
    \lambda_2 = (\frac{N\omega_a}{(\sigma_{a,v})^2})^{-1/2}
\end{align}
\begin{align}\label{eq:par3}
    \lambda_3 = (\frac{2N\omega_a}{(\sigma_{a,v})^2})^{-1/2}
\end{align}

Then, SVM is employed as the training classifier. For testing, FCID-FE will first construct the test sample set for each input image via Eq. \ref{eq:sample}. Next, the existing GMM from the training dataset is employed to encode each test image into the Fisher vector with Eq. \ref{eq:fv}. At last, FCID-FE classifies these feature vectors via the trained SVM. The algorithm of FCID-FE is summarized in Algorithm \ref{alg:fcida}.

\begin{algorithm}[t]
\caption{FCID-FE}
\raggedright
\textbf{Training Stage:}\\
\textbf{Input:} $N_{3}$ natural and fake colorized training images, the corresponding labels $L_{r,FE}$, SVM parameters\\
\textbf{Output:} $\Theta$, trained SVM classifier\\
\begin{algorithmic}[1]
\State Create samples $\Phi$ \Comment refer to Eq. \ref{eq:sample}
\State Estimate GMM model $\Theta$ from $\Phi$
\For{$i=1$ to $N_3$}
\State Encode $\Phi^i$ to $F_{FE}^i$ with $\Theta$ \Comment refer to Eq. \ref{eq:fv}
\EndFor
\State Train SVM with $F_{FE}$, $L_{r,FE}$ and SVM parameters
\end{algorithmic}
\vspace{0.3cm}
\textbf{Testing Stage:}\\
\textbf{Input:} ~~\ 
$N_{4}$ test images, $\Theta$, trained SVM classifier\\
\textbf{Output:} ~~\ 
Detection labels $L_{e,FE}$\\
\begin{algorithmic}[1]
\State Create samples $\Phi$ \Comment refer to Eq. \ref{eq:sample}
\For{$i=1$ to $N_4$}
\State Encode $\Phi^i$ to $F_{FE}^i$ with $\Theta$ \Comment refer to Eq. \ref{eq:fv}
\State Obtain $L_{e,FE}(i)$ with $F_{FE}^i$ and the trained SVM clasifier
\EndFor
\end{algorithmic}
\label{alg:fcida}
\end{algorithm}

\section{Experimental Results}\label{sec:results}

In this section, the experimental setups, evaluation measurements, databases and results are introduced accordingly.

\subsection{Setups and Measurements}

In this paper, one implementation of SVM, the LIBSVM [\ref{chang2011svm}], is employed for classification and the RBF kernel is selected. The VLFeat software [\ref{vlfeat}] is employed for GMM modeling and Fisher vector encoding.

In our experiments, both the half total error rate ($HTER$) measurement and the receiver operating characteristic (ROC) curve (with the area under the curve ($AUC$) measurement) are employed to evaluate the performances of the proposed methods. Denoting $P$, $N$, $TP$ and $TN$ as the positive samples, negative samples, true positive samples and true negative samples respectively, $HTER$ is defined as Eq. \ref{eq:hter} shows.
\begin{align}\label{eq:hter}
    HTER & = \frac{FPR+FNR}{2} \nonumber \\
         & = \frac{FP/(TN+FP)+FN/(TP+FN)}{2}
\end{align}
Note that the natural images and the fake colorized images are defined as the negative samples and the positive samples, respectively.

\subsection{Databases}

For a thorough evaluation of the proposed methods, different databases are employed/constructed for different experiments. We create the database $D1$ for parameter selection and validation by employing 10000 fake colorized images from the database ctest10k in [\ref{Larsson2016}] and their corresponding 10000 natural images from the ImageNet validation dataset [\ref{feifeili2009}]. The natural images in $D1$ include various types of images, such as animals, human, furniture and outdoor scenes.

In addition to $D1$, different databases are also prepared for assessing the performances of FCID-HIST and FCID-FE against different colorization methods. The database $D2$ consists of 2000 natural images randomly selected from the ImageNet validation dataset and their corresponding fake images, which are generated via [\ref{Larsson2016}]. The database $D3$ is constructed by randomly selecting 2000 fake colorized images from the results of [\ref{zhang2016eccv}] and 2000 corresponding natural images from the ImageNet validation dataset. The database $D4$, which contains 2000 natural images (randomly selected from the ImageNet validation dataset) and their corresponding generated fake images, is produced via employing the colorization approaches in [\ref{lizuka2016tog}]. Note that the selected natural images and their corresponding colorized images in $D2$-$D4$ are not overlapping with those in $D1$.

Similarly, databases $D5$, $D6$ and $D7$ are constructed by randomly selecting 2000 natural images from the Oxford building dataset [\ref{zisserman2007}] and generating the corresponding colorized images with [\ref{Larsson2016}], [\ref{zhang2016eccv}] and [\ref{lizuka2016tog}], respectively. Note that the real images in the Oxford building dataset [\ref{zisserman2007}] contain various content provided by "Flickr".

\begin{figure*}
  \centering
  \subfigure[]{
    \includegraphics[width=130mm]{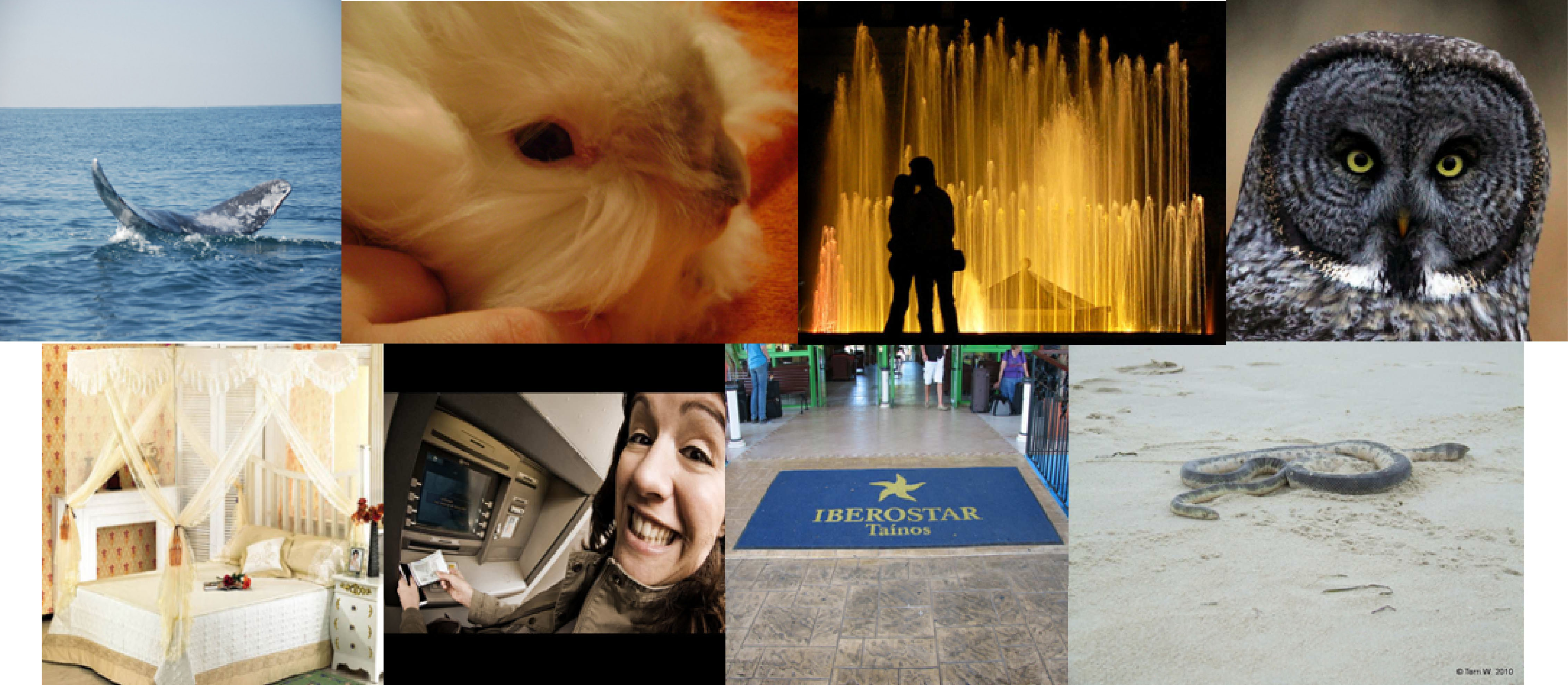}\label{fig:morerealexamples}}
  \subfigure[]{
    \includegraphics[width=130mm]{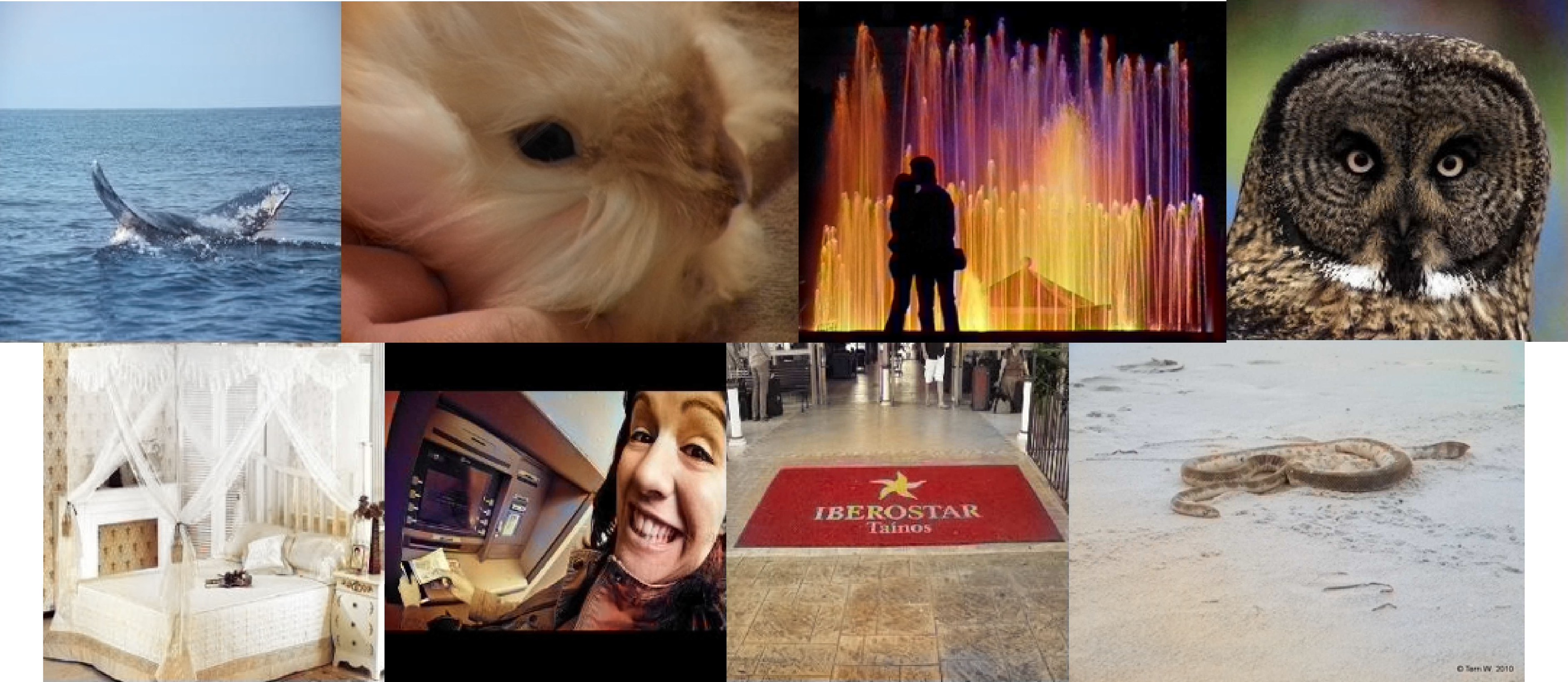}\label{fig:morefakeexamples}}
\caption{(a) Real Images. (b) Fake colorized images. \label{fig:moreexamples}}
\end{figure*}

Some examples from the databases are shown in Figs. \ref{fig:motivation} and \ref{fig:moreexamples}.

\subsection{Parameter Selection}\label{sec:parselect}

Prior to evaluating the performances of FCID-HIST and FCID-FE against different colorization approaches, the optimal parameters of the proposed methods are tuned via experiments. In the experiments, 1000 forged images and their corresponding natural images are randomly selected from database $D1$ to construct the parameter training (par-train) set, while another 1000 fake images and their corresponding natural images are selected from $D1$ to be the parameter testing (par-test) set. Note that the par-train set and the par-test set are not overlapping.

Here, two parameters, $c$ and $g$, which stand for the cost and gamma in LIBSVM, are specifically tuned here via grid search. Tables \ref{tab:svmparameterfcidb} and \ref{tab:svmparameterfcida} present the $HTER$ results with different $c$s and $g$s for FCID-HIST and FCID-FE, respectively. As shown, FCID-HIST should select $c=32$, while FCID-FE should select $c=2$ for the parameter $c$. Since there exists multiple choices for $g$, for convenience and consistency, $g=1/2$ is selected for both FCID-HIST and FCID-FE in the rest of this paper.

\begin{table*}
\begin{center}
\caption{HTER of FCID-HIST for different SVM parameter settings (in Percentage)} \label{tab:svmparameterfcidb}
\begin{tabular}{|c|c|c|c|c|c|c|c|c|c|c|c|c|c|}
  \hline
   & g=1/64 & g=1/32 & g=1/16 & g=1/8 & g=1/4 & g=1/2 & g=1 & g=2 & g=4 & g=8 & g=16 & g=32 & g=64 \\
  \hline
  c=1/64 & 24.30 & 24.30 & 24.30 & 24.30 & 24.30 & 24.30 & 24.30 & 24.05 & 23.20 & 23.40 & 23.95 & 24.60 & 26.15 \\
  c=1/32 & 24.30 & 24.30 & 24.30 & 24.30 & 24.30 & 24.30 & 24.30 & 24.05 & 23.20 & 23.40 & 23.95 & 24.60 & 26.15 \\
  c=1/16 & 24.30 & 24.30 & 24.30 & 24.30 & 24.30 & 24.30 & 24.30 & 24.05 & 23.20 & 23.40 & 23.95 & 24.60 & 26.15 \\
  c=1/8 & 24.30 & 24.30 & 24.30 & 24.30 & 24.30 & 24.30 & 24.30 & 24.05 & 23.20 & 23.40 & 23.95 & 24.60 & 26.15 \\
  c=1/4 & 24.30 & 24.30 & 24.30 & 24.30 & 24.30 & 24.30 & 24.30 & 24.05 & 23.20 & 23.40 & 23.95 & 24.60 & 26.15 \\
  c=1/2 & 24.30 & 24.30 & 24.30 & 24.30 & 24.30 & 24.30 & 24.30 & 24.05 & 23.20 & 23.40 & 23.95 & 24.60 & 26.15 \\
  c=1 & 24.30 & 24.30 & 24.30 & 24.30 & 24.30 & 24.30 & 24.30 & 24.05 & 23.20 & 23.40 & 23.95 & 24.60 & 26.15 \\
  c=2 & 23.55 & 23.55 & 23.55 & 23.55 & 23.55 & 23.55 & 23.55 & 22.65 & 22.65 & 23.05 & 23.50 & 24.45 & 26.55 \\
  c=4 & 22.90 & 22.90 & 22.90 & 22.90 & 22.90 & 22.90 & 22.90 & 22.15 & 22.20 & 22.80 & 23.90 & 25.15 & 27.80 \\
  c=8 & 22.30 & 22.30 & 22.30 & 22.30 & 22.30 & 22.30 & 22.30 & 22.20 & 22.50 & 22.85 & 23.70 & 25.95 & 28.55 \\
  c=16 & 22.00 & 22.00 & 22.00 & 22.00 & 22.00 & 22.00 & 22.00 & 21.75 & 21.90 & 23.25 & 24.40 & 26.75 & 29.10 \\
  c=32 & \textbf{21.50} & \textbf{21.50} & \textbf{21.50} & \textbf{21.50} & \textbf{21.50} & \textbf{21.50} & \textbf{21.50} & 21.65 & 22.15 & 24.20 & 24.95 & 27.75 & 30.55 \\
  c=64 & 21.65 & 21.65 & 21.65 & 21.65 & 21.65 & 21.65 & 21.65 & 22.15 & 22.50 & 24.10 & 25.75 & 28.30 & 31.00 \\
  \hline
\end{tabular}
\end{center}
\end{table*}

\begin{table*}
\begin{center}
\caption{HTER of FCID-FE for different SVM parameter settings (in Percentage)} \label{tab:svmparameterfcida}
\begin{tabular}{|c|c|c|c|c|c|c|c|c|c|c|c|c|c|}
  \hline
   & g=1/64 & g=1/32 & g=1/16 & g=1/8 & g=1/4 & g=1/2 & g=1 & g=2 & g=4 & g=8 & g=16 & g=32 & g=64 \\
  \hline
  c=1/64 & 16.90 & 16.90 & 16.90 & 16.90 & 16.90 & 16.90 & 16.90 & 19.20 & 22.25 & 27.05 & 35.05 & 58.25 & 53.20 \\
  c=1/32 & 16.90 & 16.90 & 16.90 & 16.90 & 16.90 & 16.90 & 16.90 & 19.20 & 22.25 & 27.05 & 35.05 & 58.25 & 53.20 \\
  c=1/16 & 16.90 & 16.90 & 16.90 & 16.90 & 16.90 & 16.90 & 16.90 & 19.20 & 22.25 & 27.05 & 35.05 & 58.25 & 53.20 \\
  c=1/8 & 16.90 & 16.90 & 16.90 & 16.90 & 16.90 & 16.90 & 16.90 & 19.20 & 22.25 & 27.05 & 35.05 & 58.25 & 53.20 \\
  c=1/4 & 16.90 & 16.90 & 16.90 & 16.90 & 16.90 & 16.90 & 16.90 & 19.20 & 22.25 & 27.05 & 35.05 & 58.25 & 53.20 \\
  c=1/2 & 16.90 & 16.90 & 16.90 & 16.90 & 16.90 & 16.90 & 16.90 & 19.20 & 22.25 & 27.05 & 35.05 & 58.25 & 53.20 \\
  c=1 & 16.90 & 16.90 & 16.90 & 16.90 & 16.90 & 16.90 & 16.90 & 19.20 & 22.25 & 27.05 & 35.05 & 58.25 & 53.20 \\
  c=2 & \textbf{16.65} & \textbf{16.65} & \textbf{16.65} & \textbf{16.65} & \textbf{16.65} & \textbf{16.65} & \textbf{16.65} & 19.25 & 21.60 & 26.80 & 34.70 & 58.85 & 53.70 \\
  c=4 & 17.35 & 17.35 & 17.35 & 17.35 & 17.35 & 17.35 & 17.35 & 19.15 & 21.70 & 26.80 & 34.70 & 58.85 & 53.70 \\
  c=8 & 17.45 & 17.45 & 17.45 & 17.45 & 17.45 & 17.45 & 17.45 & 19.25 & 21.65 & 26.80 & 34.70 & 58.85 & 53.70 \\
  c=16 & 17.50 & 17.50 & 17.50 & 17.50 & 17.50 & 17.50 & 17.50 & 19.60 & 21.65 & 26.80 & 34.70 & 58.85 & 53.70 \\
  c=32 & 18.25 & 18.25 & 18.25 & 18.25 & 18.25 & 18.25 & 18.25 & 19.55 & 21.65 & 26.80 & 34.70 & 58.85 & 53.70 \\
  c=64 & 18.70 & 18.70 & 18.70 & 18.70 & 18.70 & 18.70 & 18.70 & 19.55 & 21.65 & 26.80 & 34.70 & 58.85 & 53.70 \\
  \hline
\end{tabular}
\end{center}
\end{table*}

Next, we study the selection of the SVM threshold, which is important for the final classification step after the probabilities are estimated. In the test, the threshold varies from $0$ to $1$ with a step size of $0.01$. For each proposed method, a 10-fold cross threshold selection test is performed to obtain the optimal threshold by employing $D1$. Table \ref{tab:thresfcid} presents the optimal thresholds of each fold for FCID-HIST and FCID-FE. Therefore, the optimal thresholds for FCID-HIST and FCID-FE, which are calculated via averaging the optimal thresholds of each fold, are $0.455$ and $0.492$, respectively. Note that the selected thresholds for FCID-HIST and FCID-FE will be employed in the subsequent experiments.

\begin{table*}
\begin{center}
\caption{Optimal threshold selection of FCID-HIST and FCID-FE (Threshold)} \label{tab:thresfcid}
\begin{tabular}{|c|c|c|c|c|c|c|c|c|c|c|c|}
  \hline
  Method$\backslash$Fold Number & 1 & 2 & 3 & 4 & 5 & 6 & 7 & 8 & 9 & 10 \\
  \hline
  FCID-HIST & 0.46 & 0.45 & 0.46 & 0.45 & 0.46 & 0.43 & 0.46 & 0.46 & 0.45 & 0.47 \\
  FCID-FE & 0.5 & 0.51 & 0.52 & 0.43 & 0.47 & 0.54 & 0.45 & 0.5 & 0.49 & 0.51 \\
  \hline
\end{tabular}
\end{center}
\end{table*}

Since FCID-HIST exploits the histogram distributions to extract the detection features, the number of bins of the histograms $K_{c_f}, c_f={h,s,dc,bc}$ should be determined as well. Intuitively, when $K_{c_f}$ increases, part of the detection feature corresponding to the most distinctive bins may become less distinctive, while the rest of the detection feature corresponding to the total variations may capture more details and thus become more distinctive. To reveal the effects of $K_{c_f}$, the par-train and par-test sets and the SVM parameters determined above are employed. In this test, $K_{c_f}, {c_f}={h,s,dc,bc}$ is set to be from $200$ to $260$ with a step of $5$. Besides, we also include $K_{c_f}=256, {c_f}={h,s,dc,bc}$. As can be observed from Table \ref{tab:numofbins}, there exists no obvious trends when $K_{c_f}$ varies. By considering the latter results demonstrated in Section \ref{sec:crossvalid}, in which FCID-HIST gives unstable performances when the training dataset varies, we believe that $K_{c_f}$ is not a deterministic aspect for the performances of FCID-HIST. Therefore, $K_h=K_s=K_{dc}=K_{bc}$ are all set to be $200$ for convenience in this paper.


\begin{table*}
\begin{center}
\caption{The effects of $K_{c_f}$ in FCID-HIST (HTER, in Percentage)} \label{tab:numofbins}
\begin{tabular}{|c|c|c|c|c|c|c|c|c|c|c|c|c|c|c|}
  \hline
  $K_{c_f}$ & 200 & 205 & 210 & 215 & 220 & 225 & 230 & 235 & 240 & 245 & 250 & 255 & 256 & 260 \\
  \hline
  HTER & 21.50 & 21.75 & 23.00 & 20.95 & 21.05 & 21.10 & 20.80 & 20.10 & 20.95 & 21.30 & 20.15 & 19.65 & 20.90 & 19.90 \\
  \hline
\end{tabular}
\end{center}
\end{table*}

\subsection{Cross Validation}\label{sec:crossvalid}

\begin{figure}
  \centering
  \includegraphics[width=8.0cm]{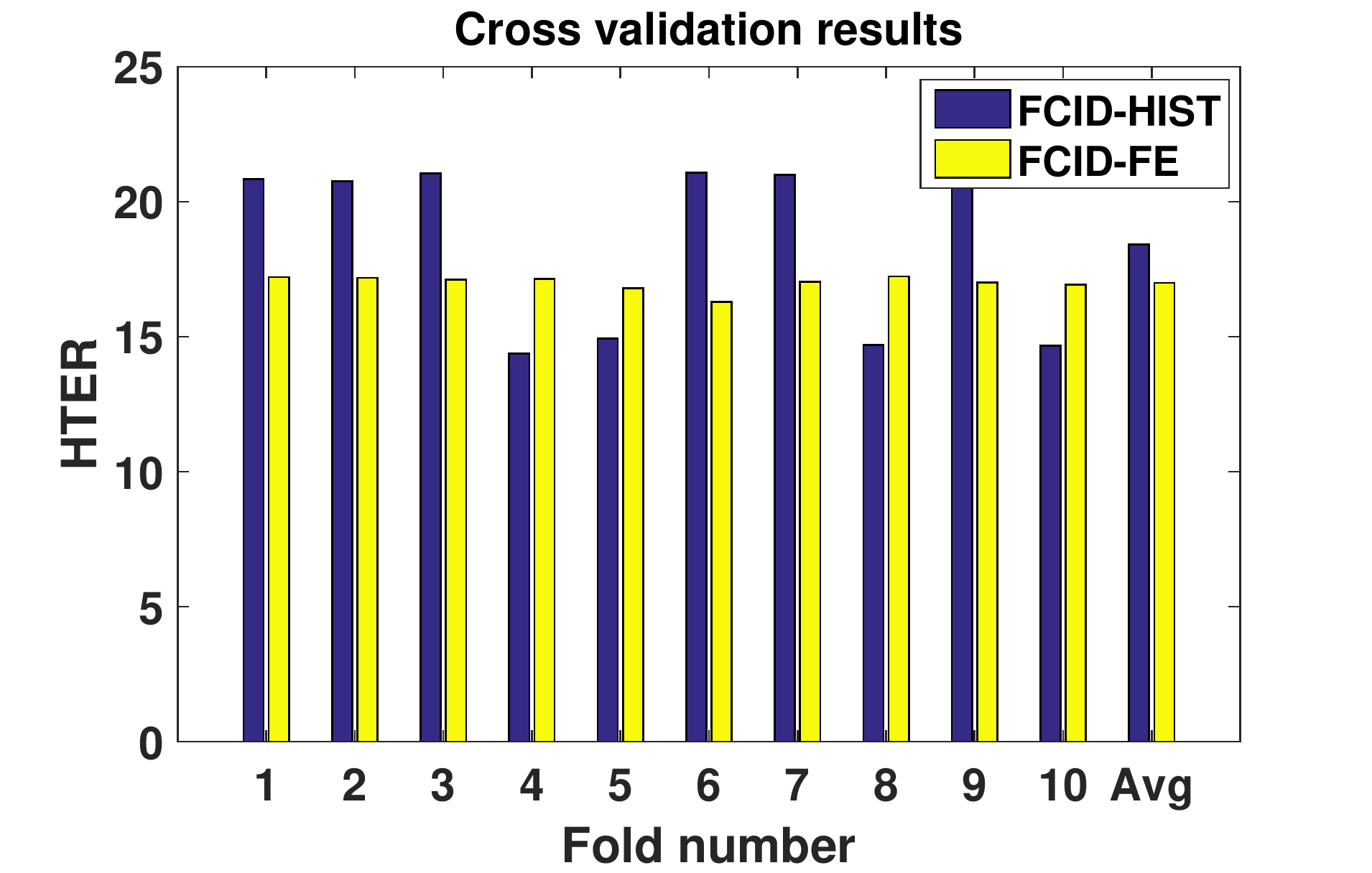}\label{fig:10foldcrossvalidationa}
\caption{FCID-HIST and FCID-FE HTER results of 10-fold cross validation. \label{fig:10foldcrossvalidation}}
\end{figure}

After the parameters are determined, the cross validations are performed on FCID-HIST and FCID-FE separately. Fig. \ref{fig:10foldcrossvalidation} presents the cross validation results of FCID-HIST and FCID-FE. As can be observed, both FCID-HIST and FCID-FE achieve a decent performance, where the average HTER of FCID-HIST is $18.423\%$ and that of FCID-FE is $16.994\%$. Clearly, FCID-FE gives a slightly better performance compared to FCID-HIST. Note that FCID-HIST gives less consistent performances because the detection feature, especially the most distinctive bins, may vary for different training set. It indicates that the extracted handcrafted features in FCID-HIST possess less robustness compared to the moments-based features in FCID-FE. The detection performances may be improved via exploring better and more consistent features in the future work.

\subsection{Performance Evaluation}

In the cross validation tests, both FCID-HIST and FCID-FE performs decently. Here, a comprehensive performance evaluation for FCID-HIST and FCID-FE is performed with six more databases $D2$, $D3$, $D4$, $D5$, $D6$ and $D7$.

Since FCID-HIST and FCID-FE construct the feasible features automatically according to the training set, the proposed methods should be capable of detecting the fake images generated by different colorization methods, as long as the colorized images exhibit the observed differences. For demonstrating the performances of the detection methods against three latest colorization approaches [\ref{Larsson2016}][\ref{zhang2016eccv}][\ref{lizuka2016tog}], each of $D2$, $D3$ and $D4$ is equally divided into a training set and a testing set.

\begin{figure*}
  \centering
  \subfigure[]{
    \includegraphics[width=5.8cm]{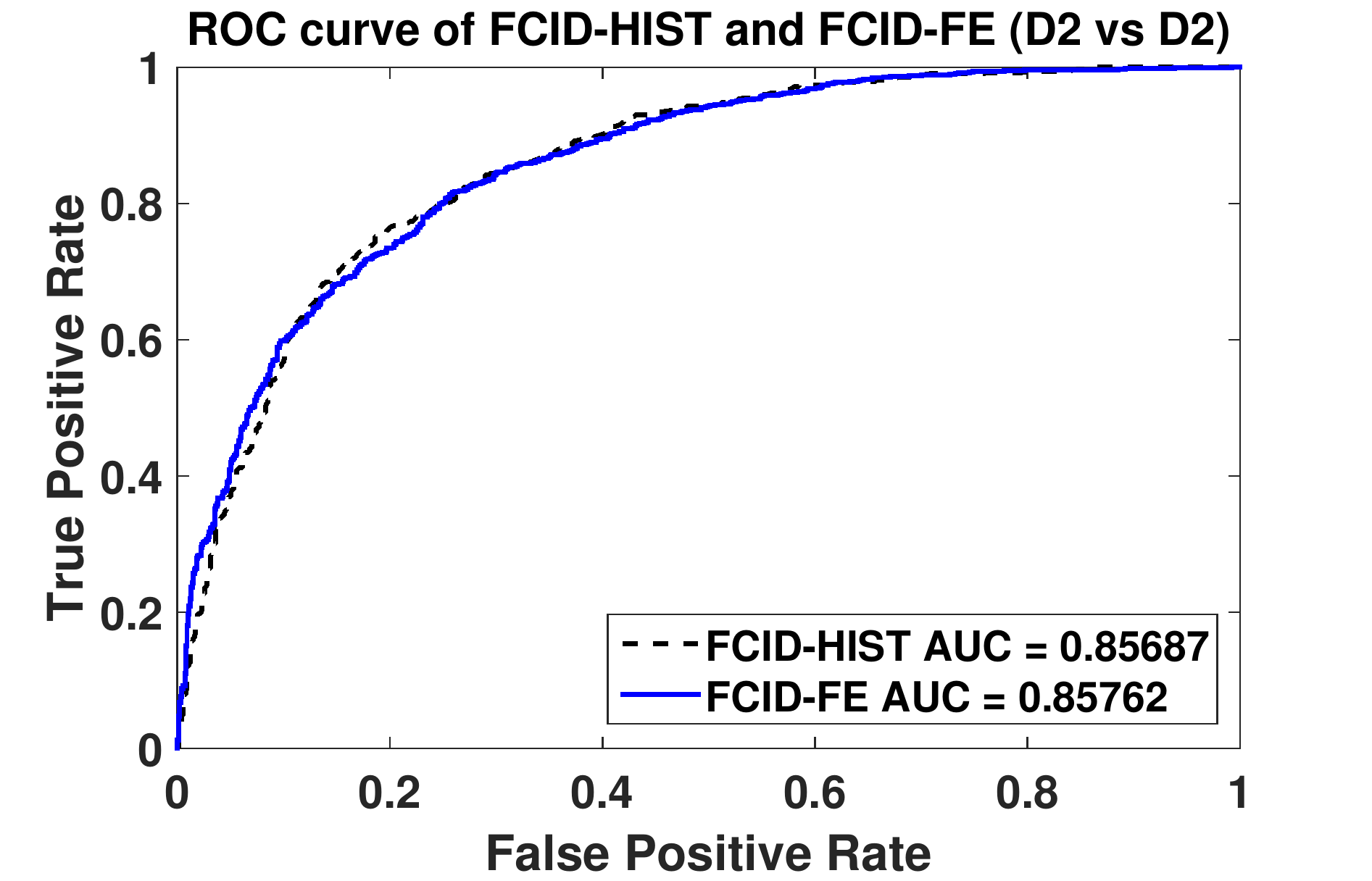}\label{fig:d2_d2}}
  \subfigure[]{
    \includegraphics[width=5.8cm]{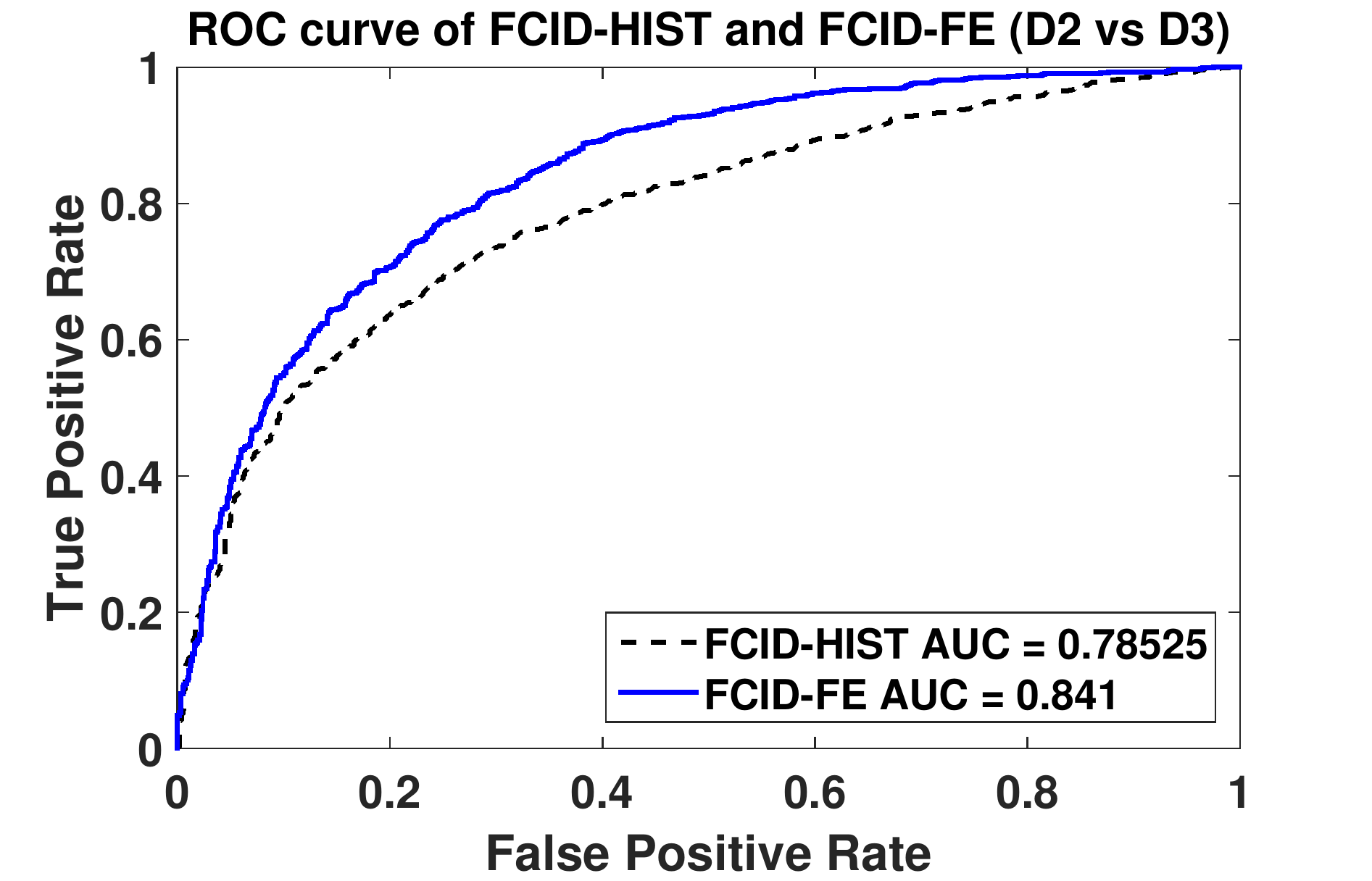}\label{fig:d2_d3}}
  \subfigure[]{
    \includegraphics[width=5.8cm]{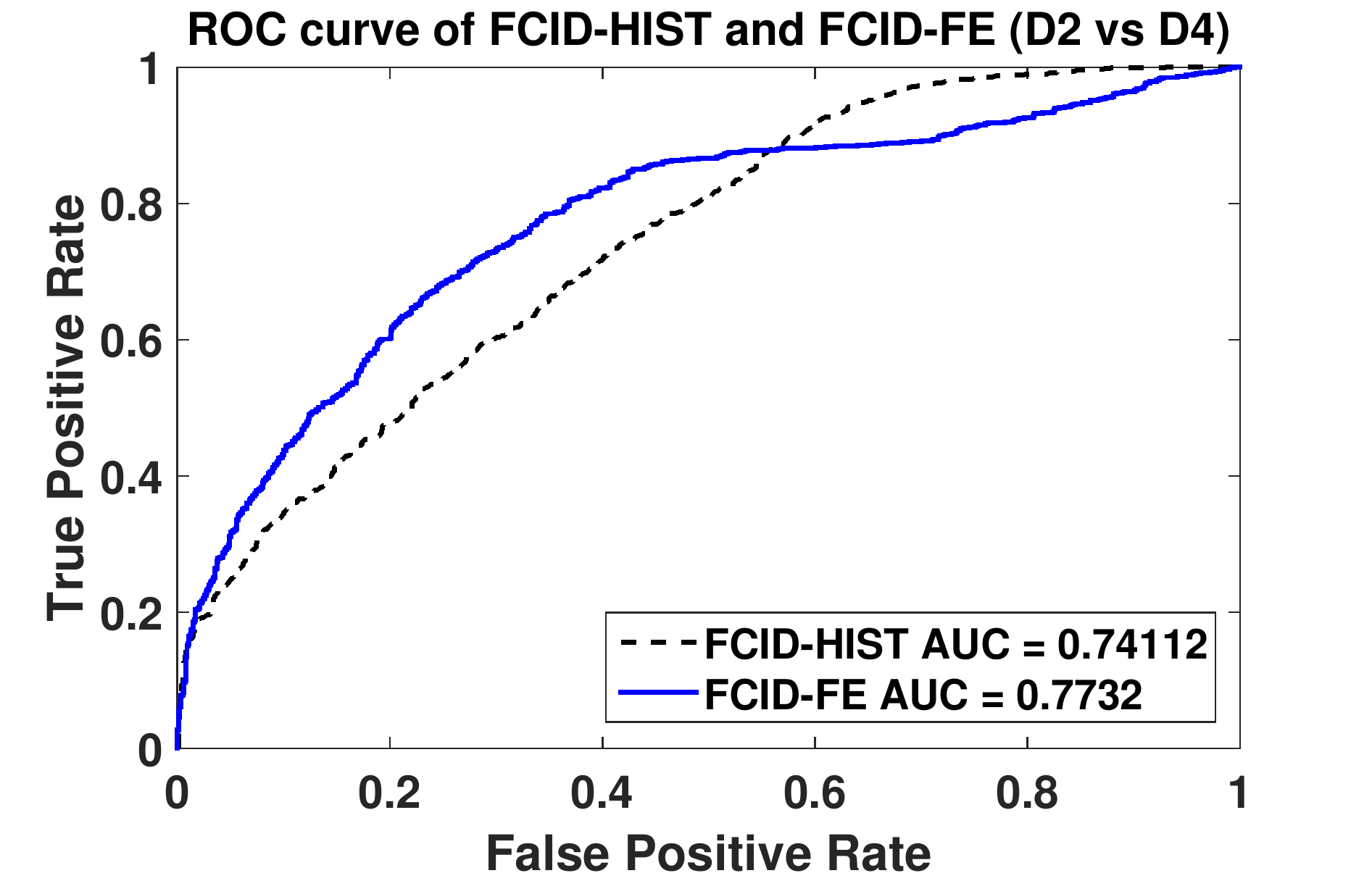}\label{fig:d2_d4}}\\
  \subfigure[]{
    \includegraphics[width=5.8cm]{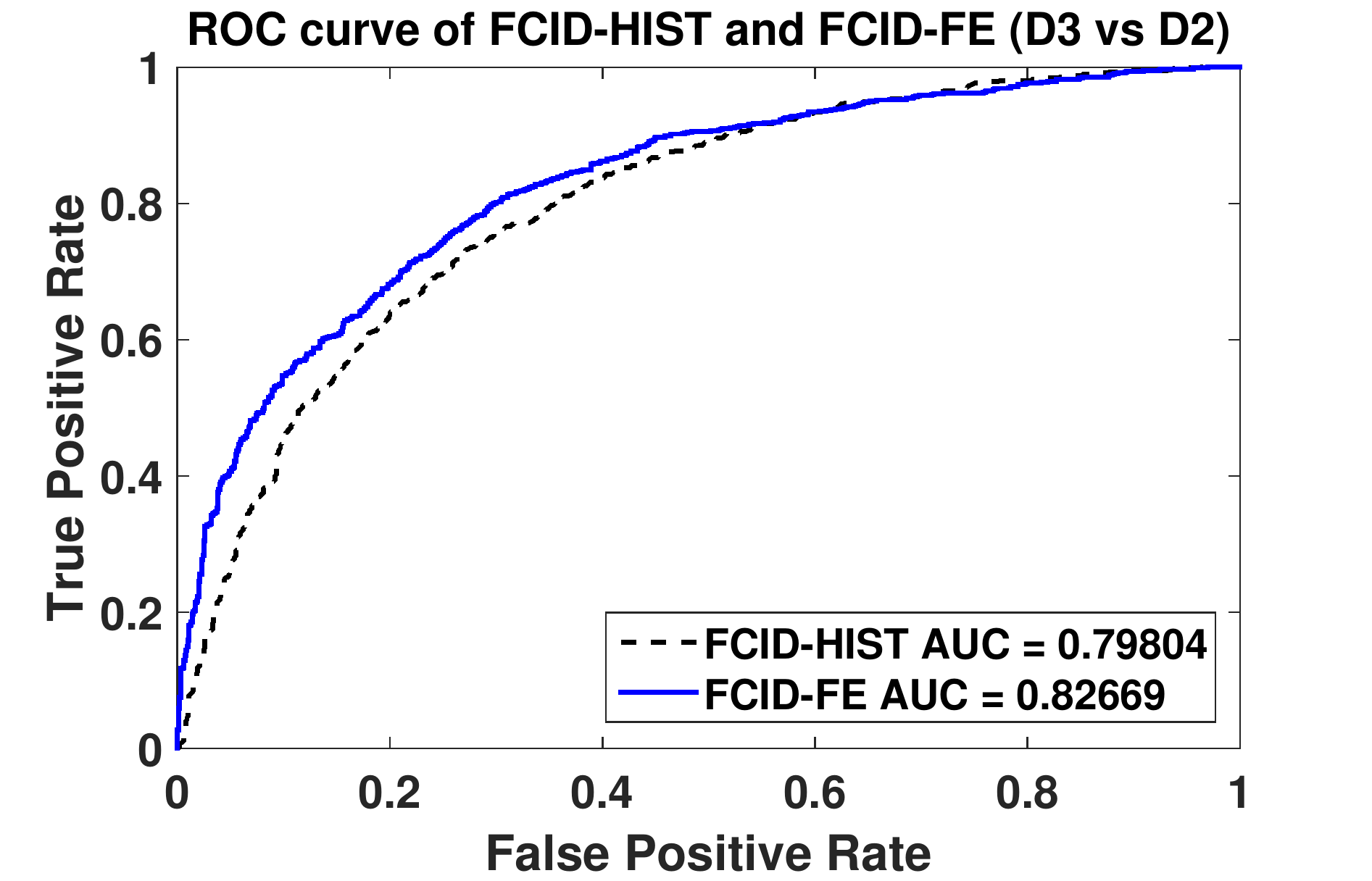}\label{fig:d3_d2}}
  \subfigure[]{
    \includegraphics[width=5.8cm]{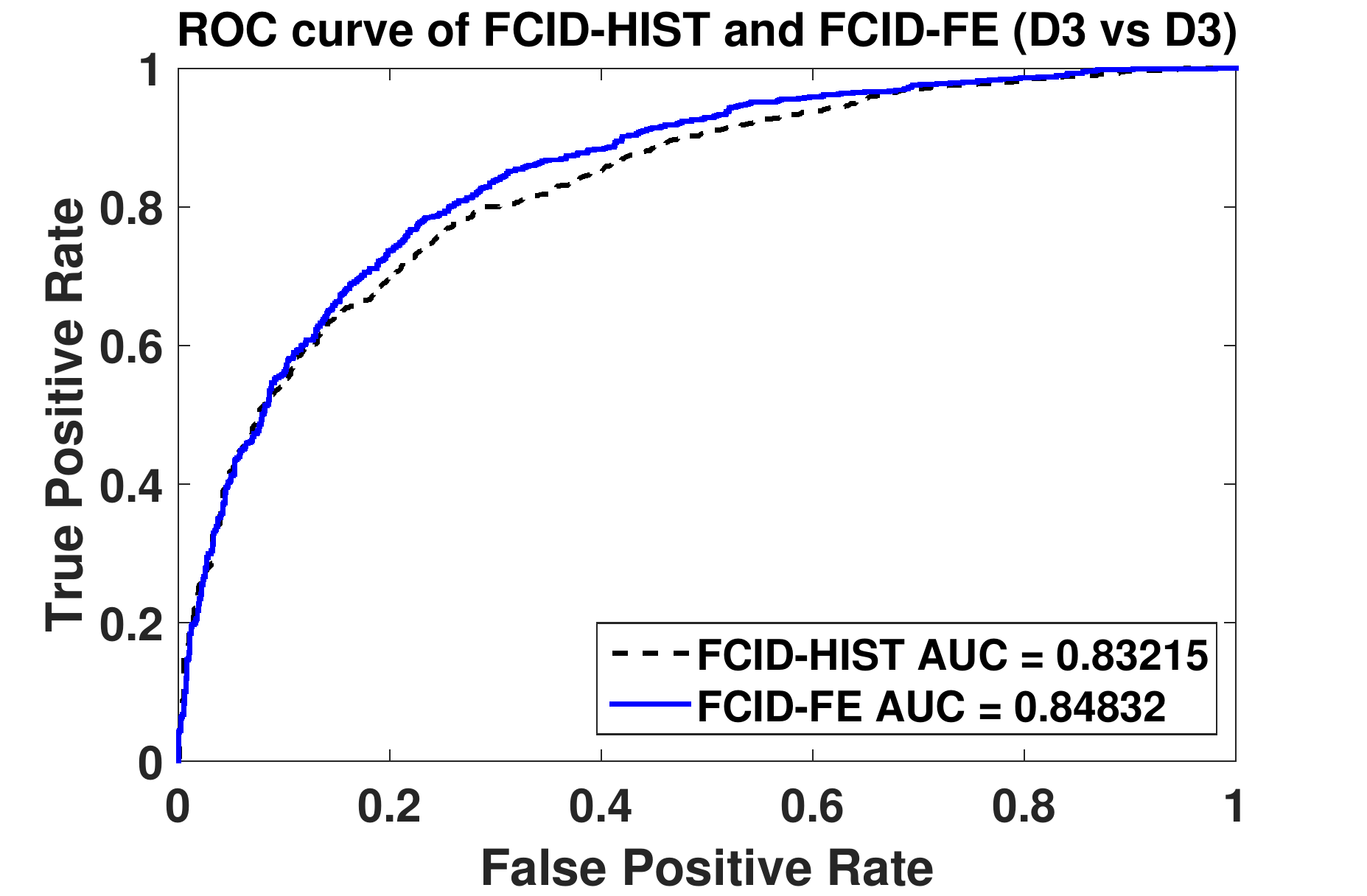}\label{fig:d3_d3}}
  \subfigure[]{
    \includegraphics[width=5.8cm]{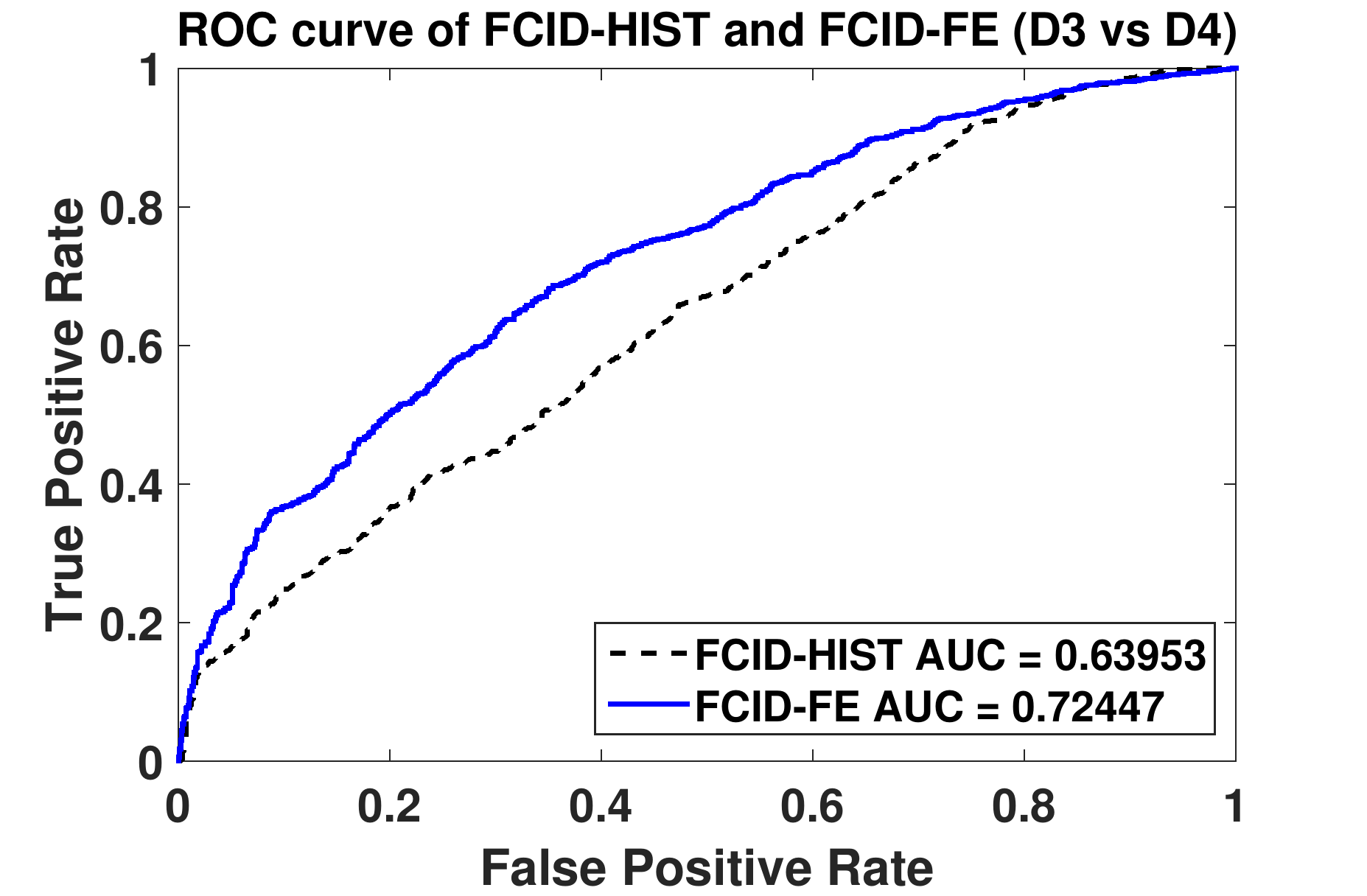}\label{fig:d3_d4}}\\
  \subfigure[]{
    \includegraphics[width=5.8cm]{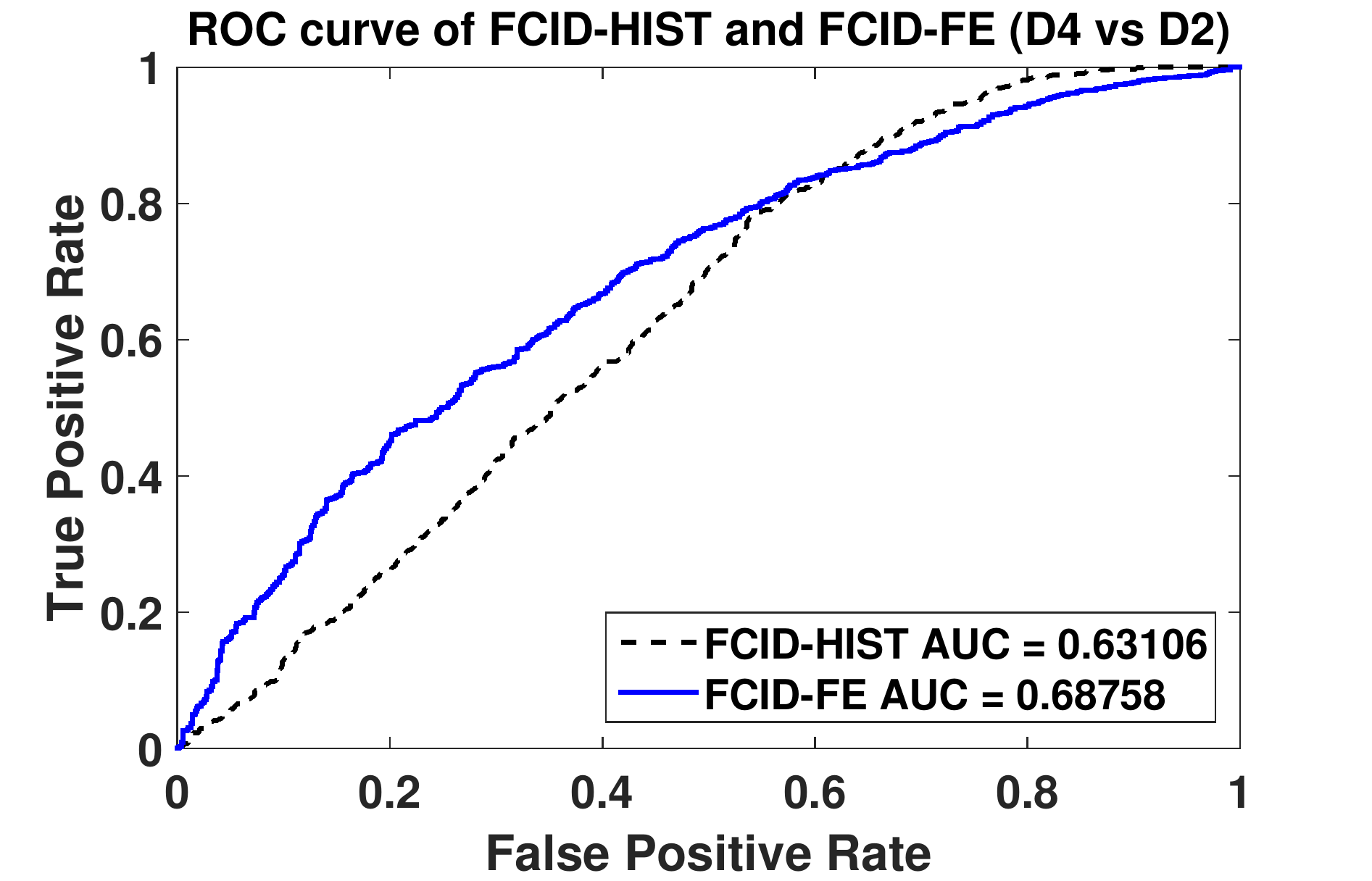}\label{fig:d4_d2}}
  \subfigure[]{
    \includegraphics[width=5.8cm]{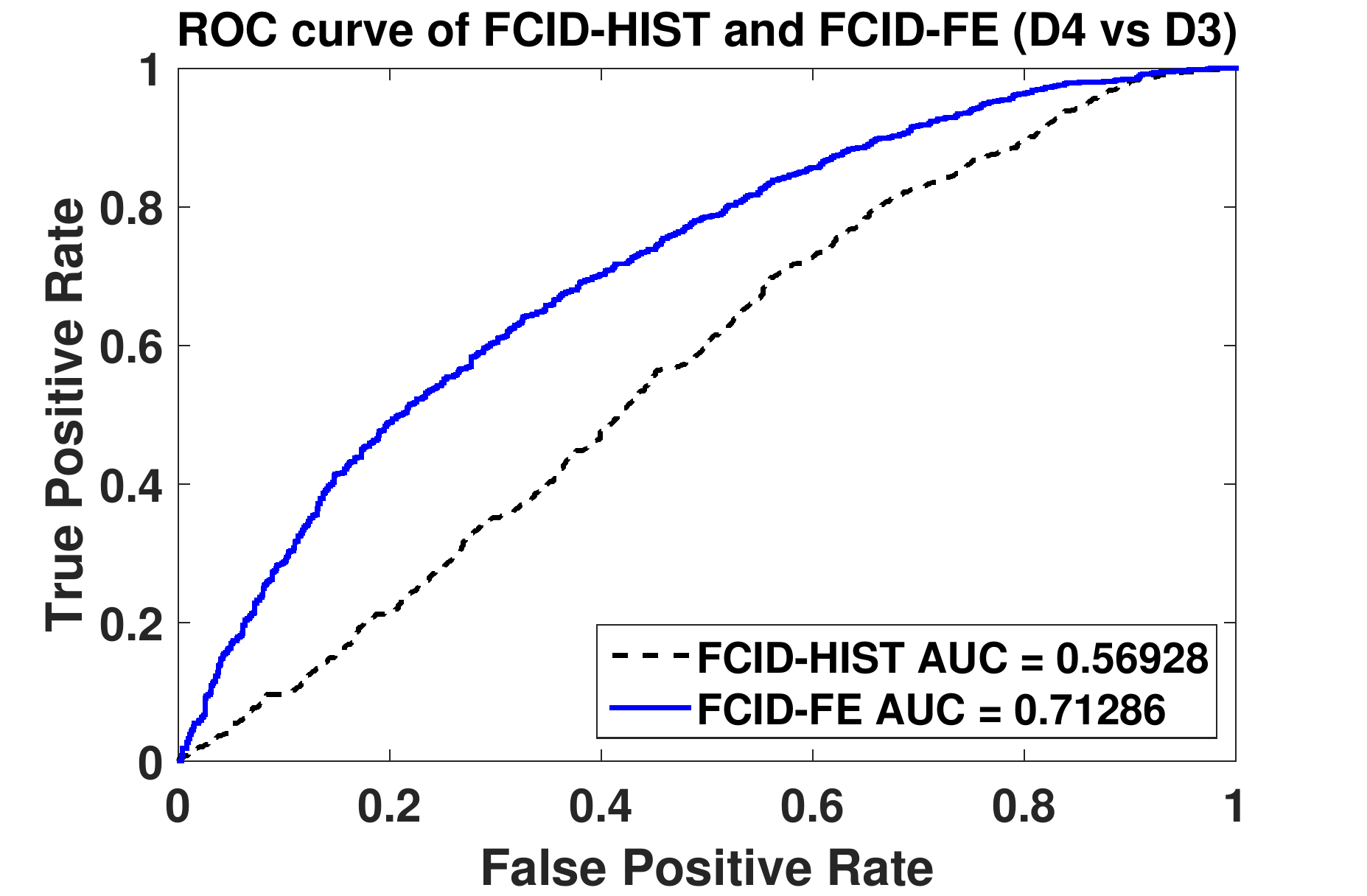}\label{fig:d4_d3}}
  \subfigure[]{
    \includegraphics[width=5.8cm]{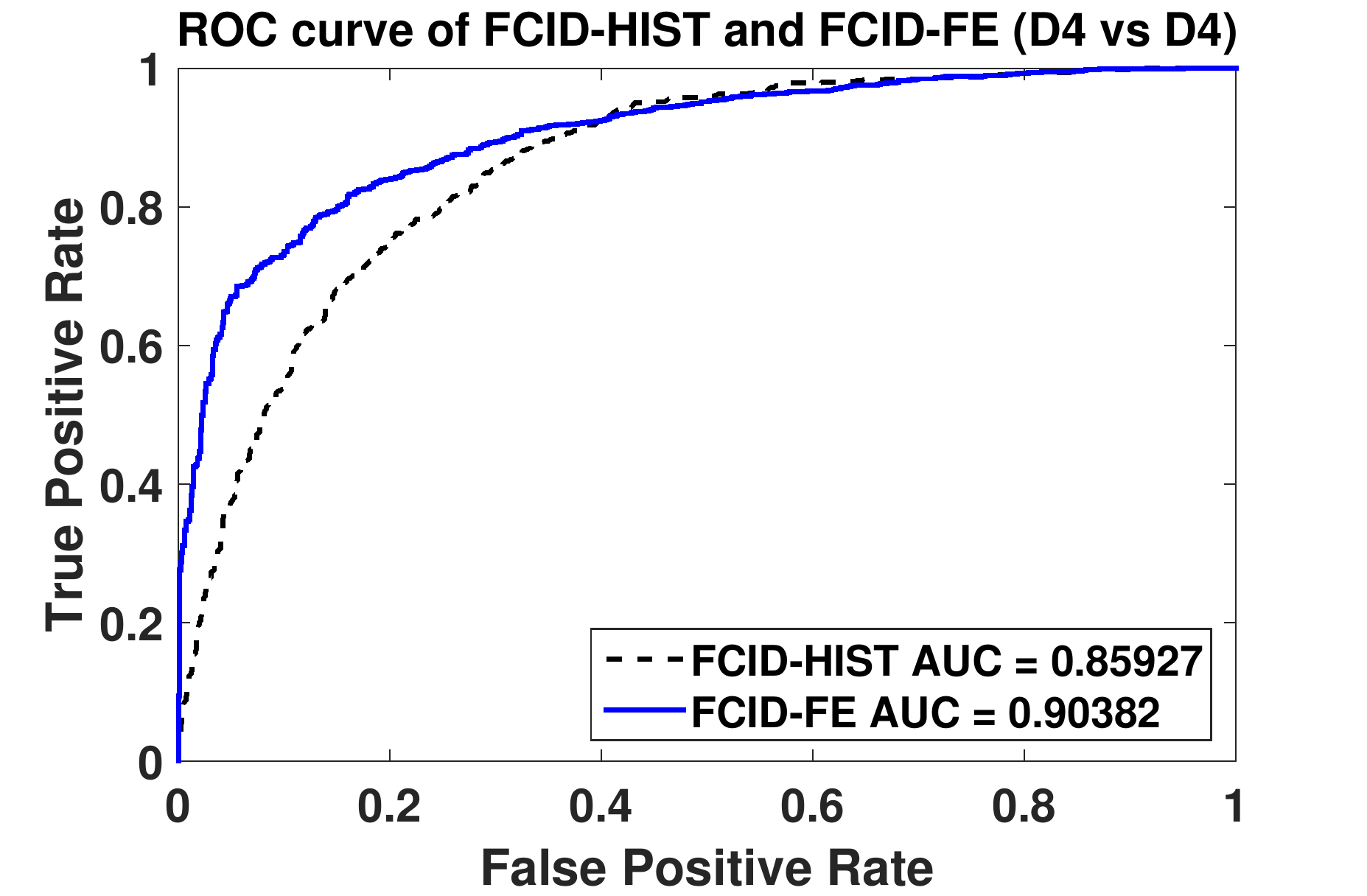}\label{fig:d4_d4}}
\caption{Detection results for the cross colorization method tests. (a) $D2$([\ref{Larsson2016}]) vs $D2$([\ref{Larsson2016}]). (b) $D2$([\ref{Larsson2016}]) vs $D3$([\ref{zhang2016eccv}]). (c) $D2$([\ref{Larsson2016}]) vs $D4$([\ref{lizuka2016tog}]). (d) $D3$([\ref{zhang2016eccv}]) vs $D2$([\ref{Larsson2016}]). (e) $D3$([\ref{zhang2016eccv}]) vs $D3$([\ref{zhang2016eccv}]). (f) $D3$([\ref{zhang2016eccv}]) vs $D4$([\ref{lizuka2016tog}]). (g) $D4$([\ref{lizuka2016tog}]) vs $D2$([\ref{Larsson2016}]). (h) $D4$([\ref{lizuka2016tog}]) vs $D3$([\ref{zhang2016eccv}]). (i) $D4$([\ref{lizuka2016tog}]) vs $D4$([\ref{lizuka2016tog}]). \label{fig:pe}}
\end{figure*}

The experiments are conducted in a manner that the training sets and testing sets may or may not be originated from the identical databases, such that 9 experiments are performed to evaluate FCID-HIST and FCID-FE. As can be observed from Tables \ref{tab:fcidbagainst3}-\ref{tab:fcidaagainst3} and Fig. \ref{fig:pe}, the proposed methods can successfully detect different fake images which are generated from different state-of-the-art colorization approaches, when the training and testing datasets are from the identical or different databases. Besides, FCID-FE gives more accurate detection results compared to FCID-HIST in most situations. Compared to Figs. \ref{fig:d2_d2}, \ref{fig:d3_d3} and \ref{fig:d4_d4}, performance decreases when the training and testing datasets are from different databases, especially for FCID-HIST. These drops reveal that FCID-FE, which gives more consistent performances, models the statistical information of the images better compared to FCID-HIST.

\begin{table}
\begin{center}
\caption{HTER of FCID-HIST for different databases (in Percentage)} \label{tab:fcidbagainst3}
\begin{tabular}{|c|c|c|c|}
  \hline
  Training$\backslash$Testing & $D2$([\ref{Larsson2016}]) & $D3$([\ref{zhang2016eccv}]) & $D4$([\ref{lizuka2016tog}]) \\
  \hline
  $D2$([\ref{Larsson2016}]) & 22.50 & 28.00 & 33.95 \\
  $D3$([\ref{zhang2016eccv}]) & 26.95 & 24.45 & 41.85 \\
  $D4$([\ref{lizuka2016tog}]) & 38.15 & 43.55 & 22.35 \\
  \hline
\end{tabular}
\end{center}
\end{table}

\begin{table}
\begin{center}
\caption{HTER of FCID-FE for different databases (in Percentage)} \label{tab:fcidaagainst3}
\begin{tabular}{|c|c|c|c|}
  \hline
  Training$\backslash$Testing & $D2$([\ref{Larsson2016}]) & $D3$([\ref{zhang2016eccv}]) & $D4$([\ref{lizuka2016tog}]) \\
  \hline
  $D2$([\ref{Larsson2016}]) & 22.30 & 23.65 & 31.70 \\
  $D3$([\ref{zhang2016eccv}]) & 25.10 & 22.85 & 34.25 \\
  $D4$([\ref{lizuka2016tog}]) & 38.50 & 36.15 & 17.30 \\
  \hline
\end{tabular}
\end{center}
\end{table}

Next, the cross dataset tests are performed. The natural images in $D2$, $D3$ and $D4$, originating from the ImageNet validation dataset [\ref{feifeili2009}], and images in the $D5$, $D6$ and $D7$, originating from the Oxford building dataset [\ref{zisserman2007}], are employed to perform the cross dataset tests.

\begin{figure*}
  \centering
  \subfigure[]{
    \includegraphics[width=5.8cm]{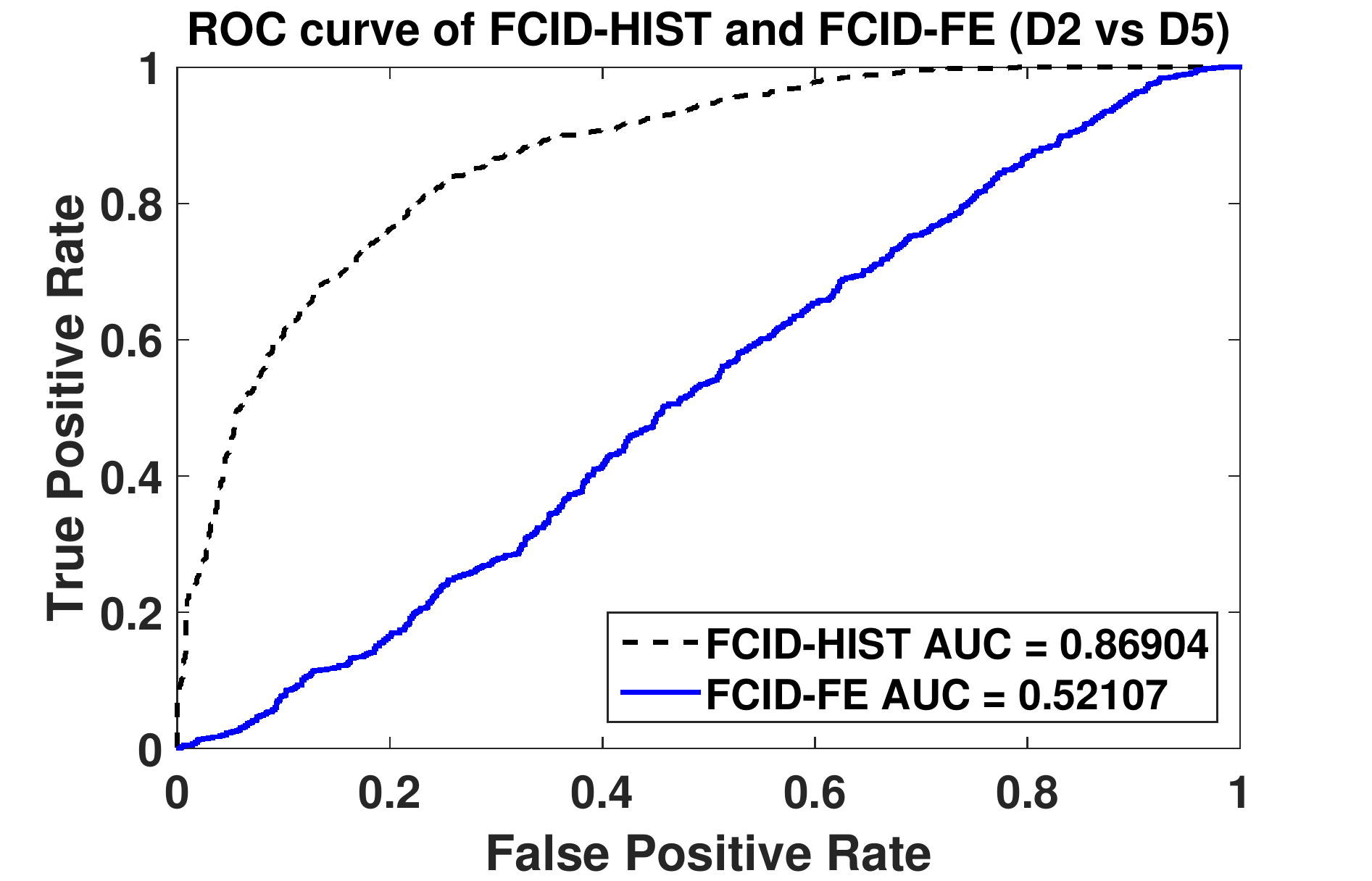}\label{fig:od2_nd2}}
  \subfigure[]{
    \includegraphics[width=5.8cm]{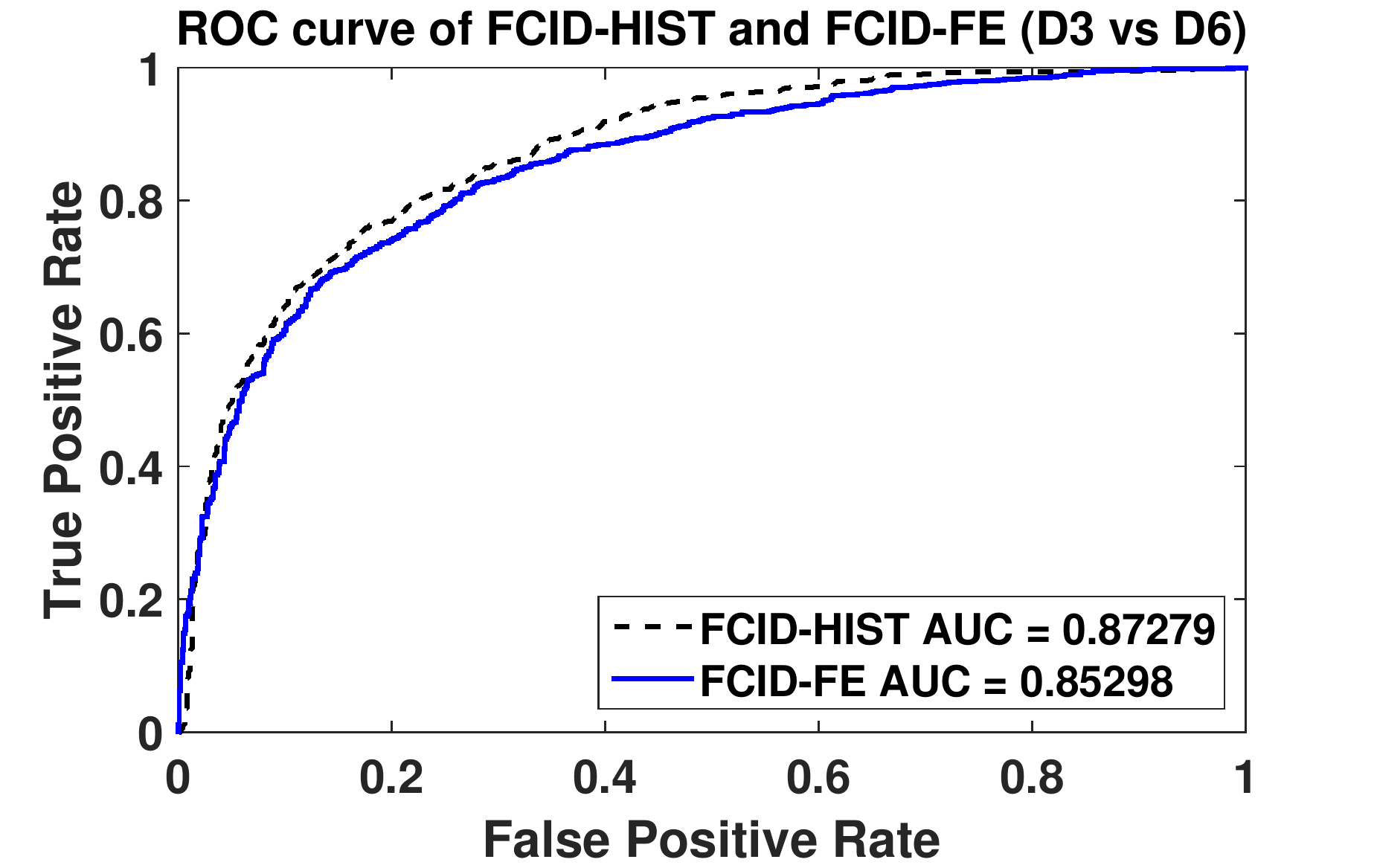}\label{fig:od3_nd3}}
  \subfigure[]{
    \includegraphics[width=5.8cm]{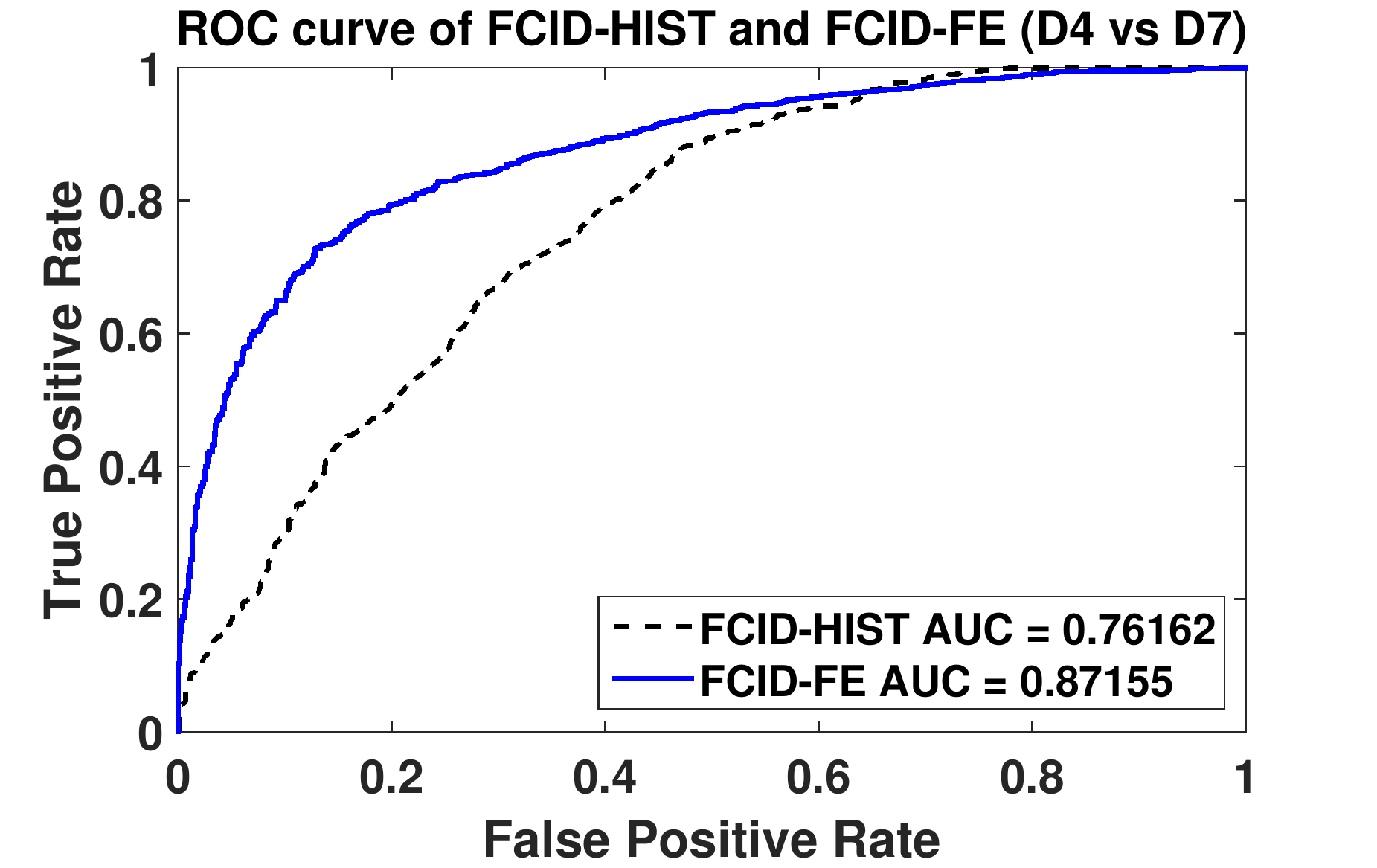}\label{fig:od4_nd4}}\\
  \subfigure[]{
    \includegraphics[width=5.8cm]{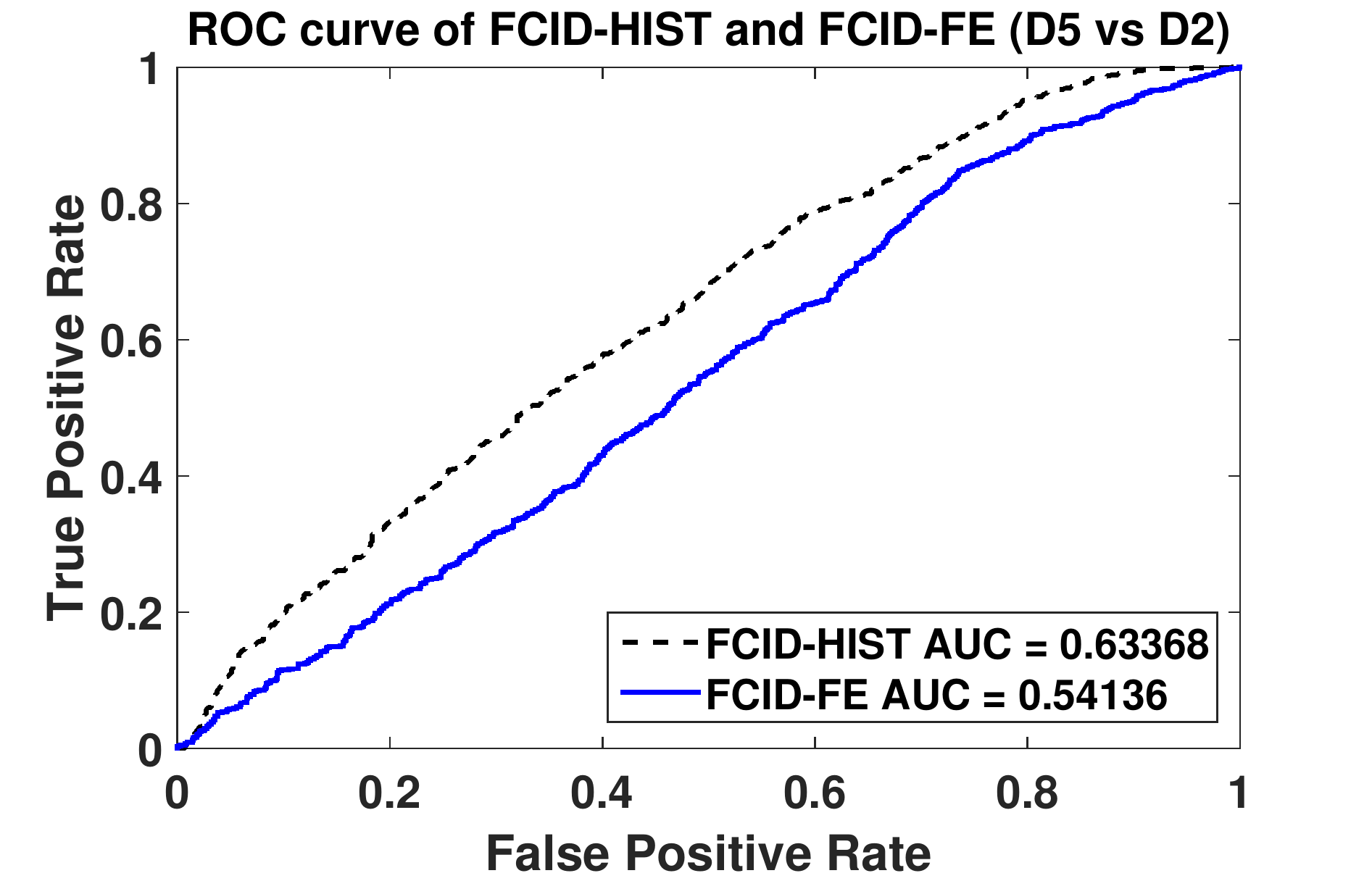}\label{fig:nd2_od2}}
  \subfigure[]{
    \includegraphics[width=5.8cm]{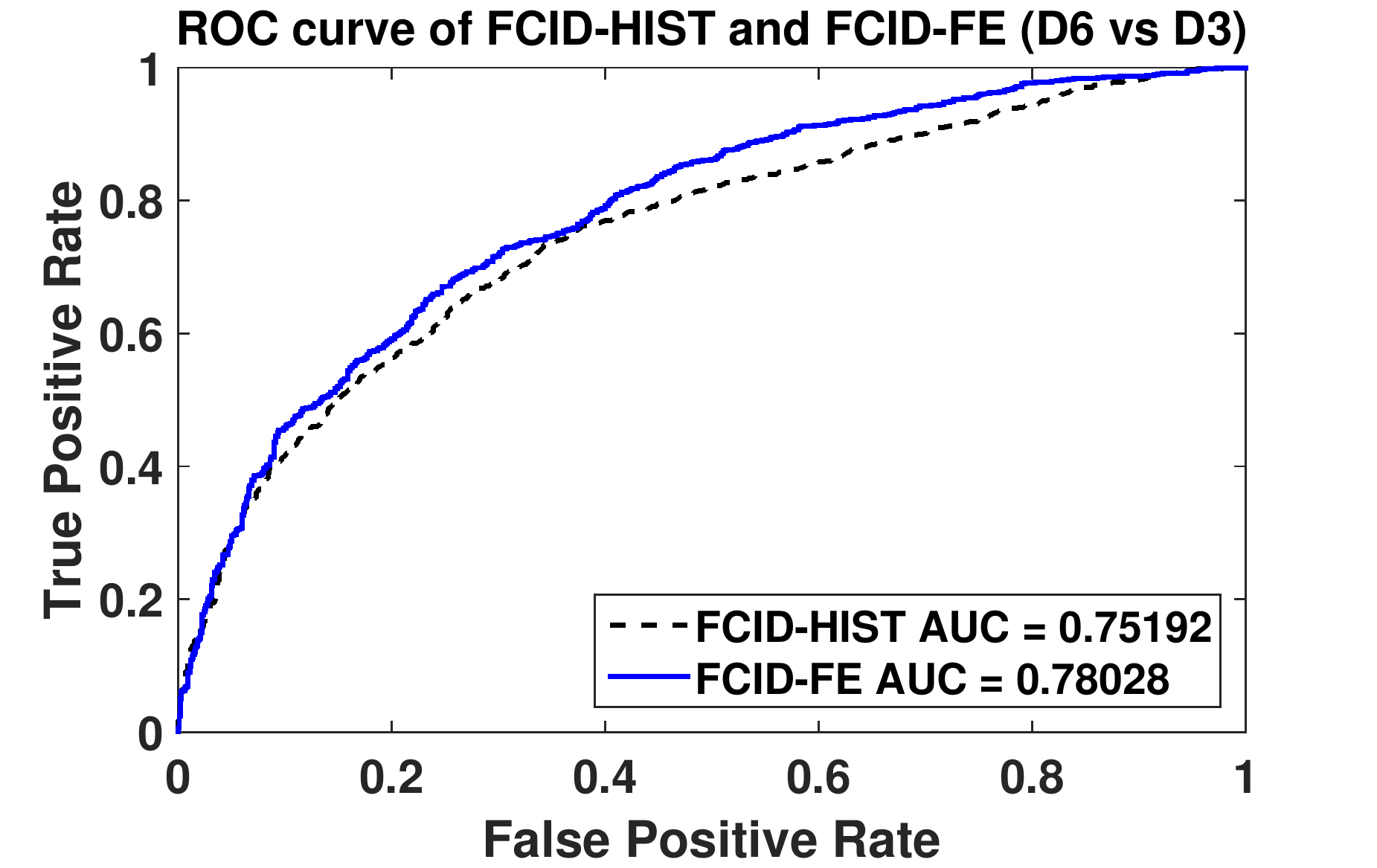}\label{fig:nd3_od3}}
  \subfigure[]{
    \includegraphics[width=5.8cm]{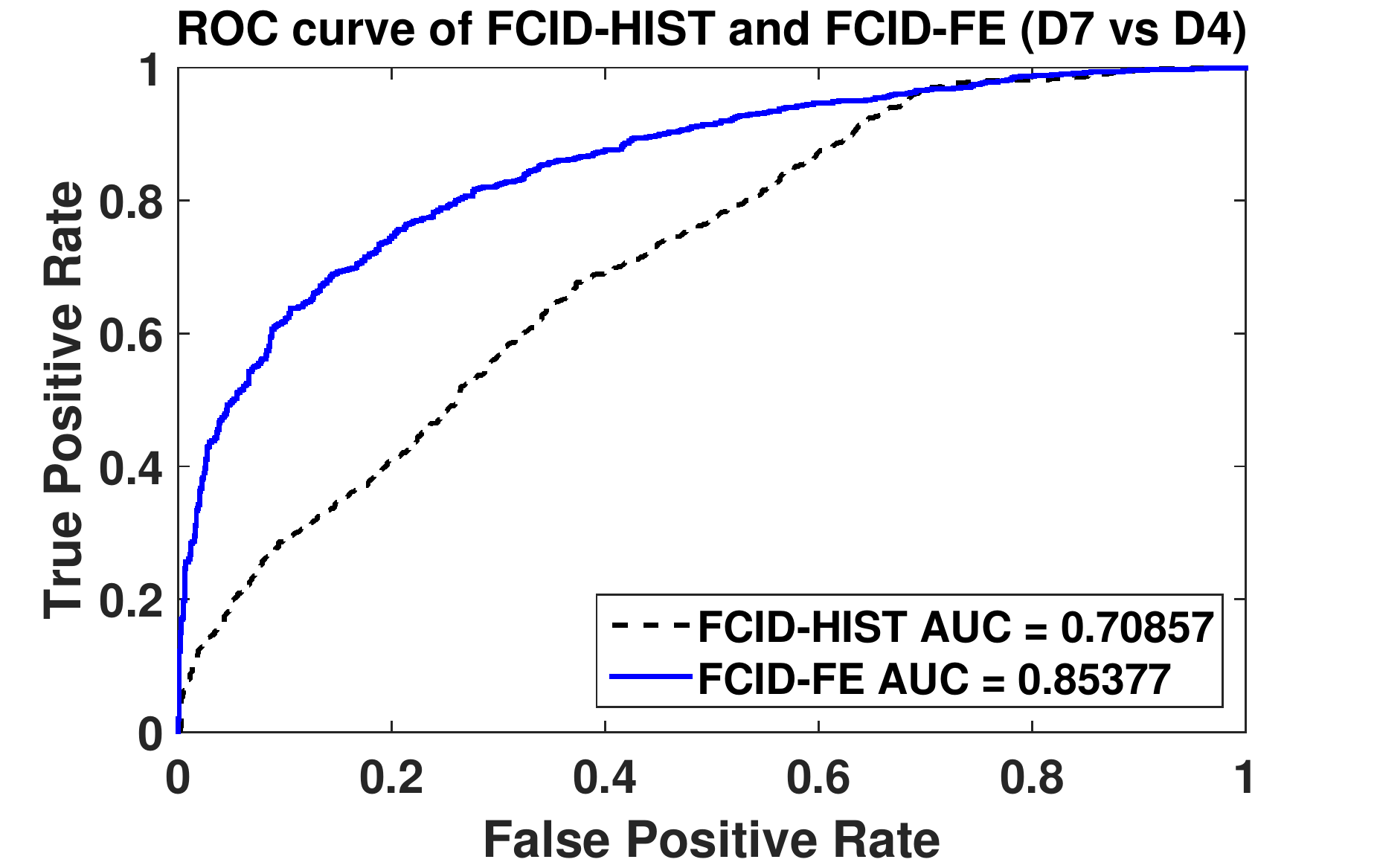}\label{fig:nd4_od4}}
\caption{Detection results with different training vs testing sets. (a) $D2$([\ref{Larsson2016}]) vs $D5$([\ref{Larsson2016}]). (b) $D3$([\ref{zhang2016eccv}]) vs $D6$([\ref{zhang2016eccv}]). (c) $D4$([\ref{lizuka2016tog}]) vs $D7$([\ref{lizuka2016tog}]). (d) $D5$([\ref{Larsson2016}]) vs $D2$([\ref{Larsson2016}]). (e) $D6$([\ref{zhang2016eccv}]) vs $D3$([\ref{zhang2016eccv}]). (f) $D7$([\ref{lizuka2016tog}]) vs $D4$([\ref{lizuka2016tog}]). \label{fig:pe_cross}}
\end{figure*}

Similar to $D2$, $D3$ and $D4$, $D5$, $D6$ and $D7$ are all equally divided into training and testing sets. By pairing the databases in which the colorized images are generated from the same colorization method, three database pairs, $D2$ and $D5$, $D3$ and $D6$, $D4$ and $D7$, are obtained. For each pair of databases, the cross-dataset tests are performed by employing one database's training set and the other one's testing set, and vice versa. The experimental results of the cross dataset tests are introduced in Tables \ref{tab:fcidbcrossdb31}-\ref{tab:fcidacrossdb31} and Fig. \ref{fig:pe_cross}. As shown, although the performance somewhat decreases, both methods still successfully differentiates between the colorized and natural images, and FCID-HIST again gives less stable performances compared to FCID-FE, with the exception of the $D2$ and $D5$ pair. The unsatisfactory performances for the $D2$ and $D5$ pair may be due to the different image content in different image datasets ($D2$ from ImageNet dataset and $D5$ from Oxford building dataset), which induces different statistical distributions. Since the proposed methods, especially FCID-FE, rely on extracting the detection features from the entire distributions, the classifier, which is trained by one of $D2$ and $D5$, may fail to correctly classify certain images in the other one.

\begin{table}
\begin{center}
\caption{HTER of FCID-HIST for cross-dataset tests (Training vs. Testing)} \label{tab:fcidbcrossdb31}
\begin{tabular}{|c|c|c|}
  \hline
  $D2$([\ref{Larsson2016}]) vs. $D5$([\ref{Larsson2016}]) & $D3$([\ref{zhang2016eccv}]) vs. $D6$([\ref{zhang2016eccv}]) & $D4$([\ref{lizuka2016tog}]) vs. $D7$([\ref{lizuka2016tog}]) \\
  \hline
  22.85 & 21.50 & 30.95 \\
  \hline
  $D5$([\ref{Larsson2016}]) vs. $D2$([\ref{Larsson2016}]) & $D6$([\ref{zhang2016eccv}]) vs. $D3$([\ref{zhang2016eccv}]) & $D7$([\ref{lizuka2016tog}]) vs. $D4$([\ref{lizuka2016tog}]) \\
  \hline
  43.45 & 30.75 & 36.60 \\
  \hline
\end{tabular}
\end{center}
\end{table}

\begin{table}
\begin{center}
\caption{HTER of FCID-FE for cross-dataset tests (Training vs. Testing)} \label{tab:fcidacrossdb31}
\begin{tabular}{|c|c|c|}
  \hline
  $D2$([\ref{Larsson2016}]) vs. $D5$([\ref{Larsson2016}]) & $D3$([\ref{zhang2016eccv}]) vs. $D6$([\ref{zhang2016eccv}]) & $D4$([\ref{lizuka2016tog}]) vs. $D7$([\ref{lizuka2016tog}]) \\
  \hline
  51.40 & 22.70 & 20.20 \\
  \hline
  $D5$([\ref{Larsson2016}]) vs. $D2$([\ref{Larsson2016}]) & $D6$([\ref{zhang2016eccv}]) vs. $D3$([\ref{zhang2016eccv}]) & $D7$([\ref{lizuka2016tog}]) vs. $D4$([\ref{lizuka2016tog}]) \\
  \hline
  49.80 & 30.25 & 23.15 \\
  \hline
\end{tabular}
\end{center}
\end{table}

In summary, these results indicate that colorization induces statistical differences in the hue, saturation, dark and bright channels, and demonstrate the robustness of our proposed methods against different colorization methods and across different datasets.

\section{Conclusion and Discussion}\label{sec:conclusion}

In this paper, we aimed to address a new problem in the field of fake image detection: fake colorized image detection. We observed that fake colorized images and their corresponding natural images possess statistical differences in the hue, saturation, dark and bright channels. We proposed two simple yet effective schemes, FCID-HIST and FCID-FE, to resolve this detection problem. FCID-HIST exploits the most distinctive bins and total variations of the normalized histogram distributions and creates features for detection, while FCID-FE models the data samples with GMM and creates Fisher vectors for better utilizing the statistical differences. We evaluate the performances of the proposed methods by selecting parameters for FCID-HIST and FCID-FE and detecting different fake images generated by state-of-the-art colorization approaches. The results demonstrate that both FCID-HIST and FCID-FE perform decently against different colorization approaches and FCID-FE gives more consistent and superior performances compared to FCID-HIST in most of the tests.

Although the proposed FCID-HIST and FCID-FE give decent performances in the experiments, this paper is only a preliminary investigation, and there are many directions for future work that require further exploration. As our results indicate, the performance of our current methods sometimes degrades obviously when the training images and the testing images are generated from different colorization methods or different datasets, thus blind fake colorized image detection features and methods may be developed in the future by studying the common characteristics of different colorization methods. Moreover, better feature encoding approaches can be considered for improving performance, as well as the optimization of the detection features and parameters to improve the custom features constructed in this study.


\begin{thebibliography}{1}

\bibitem{}\label{farid2009}
H. Farid, ``Exposing digital forgeries from JPEG ghosts,'' \emph{IEEE Trans. Inf. Forensics and Security}, vol. 4, no. 1, pp. 154-160, 2009.

\bibitem{}\label{Li2015}
J. Li, X. Li, B. Yang and X. Sun, ``Segmentation-Based Image Copy-Move Forgery Detection Scheme,'' \emph{IEEE Trans. Inf. Forensics and Security}, vol. 10, no. 3, pp. 507-518, 2015.

\bibitem{}\label{cao2014}
G. Cao, Y. Zhao, R. Ni and X. Li, ``Contrast Enhancement-Based Forensics in Digital Images,'' \emph{IEEE Trans. Inf. Forensics and Security}, vol. 9, no. 3, pp. 515-525, 2014.

\bibitem{}\label{Larsson2016}
G. Larsson, M. Maire and G. Shakhnarovich, ``Learning representations for automatic colorization,'' \emph{in Proc. European Conf. Comp. Vision (ECCV)}, pp. 577-593, 2016.

\bibitem{}\label{goodfellow2014gan}
I.J. Goodfellow, J. Pouget-Abadie, M. Mirza, B. Xu, D. Warde-Farley, S. Ozair, A. Courville and Y. Bengio, ``Generative adversarial nets,'' \emph{in Procs. Advances in Neural Inf. Process. Systems (NIPS)}, pp. 2672-2680, 2014.

\bibitem{}\label{huang2016rdh}
F. Huang, X. Qu, H.J. Kim and J. Huang, ``Reversible data hiding in JPEG images,'' \emph{IEEE Trans. Circuits and Systems for Video Technology}, vol. 26, no. 9, pp. 1610-1621, 2016.

\bibitem{}\label{yin2016iwdw}
J. Yin, R. Wang, Y. Guo and F. Liu, ``An adaptive reversible data hiding scheme for JPEG images,'' \emph{in Proc. Int. Workshop on Digital-Forensics and Watermarking (IWDW)}, pp. 456-469, 2016

\bibitem{}\label{wang2017multiw}
J. Wang, S. Lian and Y.-Q. Shi, "Hybrid multiplicative multi-watermarking in DWT domain", \emph{Multidimensional Systems and Signal Process.}, vol. 28, no. 2, pp. 617¨C636, 2017.

\bibitem{}\label{yang2017ecp}
Y. Yang, W. Ren, Y. Guo, R. Wang and X. Cao, ``Image deblurring via extreme channels prior,'' \emph{in Procs. IEEE Int. Conf. Comp. Vision and Pattern Recognition (CVPR)}, 2017, Accepted.

\bibitem{}\label{far2005}
J. Farquhar, S. Szedmak, H. Meng and J. Shawe-Taylor, ``Improving "bag-of-keypoints" image categorization,'' \emph{Technical report, University of Southampton}, 2005.

\bibitem{}\label{perronnin2007fv}
F. Perronnin and C. Dance, ``Fisher kernels on vocabularies for image categorization,'' \emph{in Proc. IEEE Int. Conf. Comp. Vision and Pattern Recognition (CVPR)}, pp. 1-8, 2007.

\bibitem{}\label{qureshi2015survey}
M.A. Qureshi and M. Deriche, ``A bibliography of pixel-based blind image forgery detection techniques,'' \emph{Signal Process.: Image Commun.}, vol. 39, pp. 46-74, 2015.

\bibitem{}\label{fridrich2003}
J. Fridrich, D. Soukal and J. Lukas, ``Detection of copy-move forgery in digital images,'' \emph{in Proc. Digital Forensic Research Workshop}, 2003.

\bibitem{}\label{Yu2010}
W. Li and N. Yu, ``Rotation robust detection of copy-move forgery,'' \emph{in Proc. IEEE Int. Conf. Image Process. (ICIP)}, pp. 2113-2116, 2010.

\bibitem{}\label{kirchner2013}
S.-J. Ryu, M. Kirchner, M.-J. Lee, H.-K. Lee, ``Rotation invariant localization of duplicated image regions based on Zernike moments,'' \emph{IEEE Trans. Inf. Forensics and Security}, vol. 8, no. 8, pp. 1355-1370, 2013.

\bibitem{}\label{Amerini2011}
I. Amerini, L. Ballan, R. Caldelli, A. Del Bimbo and G. Serra, ``A SIFT-Based Forensic Method for Copy¨CMove Attack Detection and Transformation Recovery,'' \emph{IEEE Trans. Inf. Forensics and Security}, vol. 6, no. 3, pp. 1099-1110, 2011.

\bibitem{}\label{lowe2004sift}
D.G. Lowe, ``Distinctive image features from scale-invariant keypoints,'' \emph{Int. J. Comp. Vision}, vol. 60, no. 2, pp. 91-110, 2004.

\bibitem{}\label{Liu2014}
L. Liu, R. Ni, Y. Zhao and S. Li, ``Improved SIFT-Based Copy-Move Detection Using BFSN Clustering and CFA Features,'' \emph{in Proc. IEEE Int. Conf. Intelligent Inf. Hiding and Multimedia Signal Process. (IIHMSP)}, pp. 626-629, 2014.

\bibitem{}\label{Jiantao2016}
Y. Li and J. Zhou, ``Image copy-move forgery detection using hierarchical feature point matching,'' \emph{in Proc. Asia-Pacific Signal and Inf. Process. Association Annual Summit and Conf. (APSIPA ASC)}, pp. 1-4, 2016.

\bibitem{}\label{bianchi2012}
T. Bianchi and A. Piva, ``Image forgery localization via block-grained analysis of JPEG artifacts,'' \emph{IEEE Trans. Inf. Forensics and Security}, vol. 7, no. 3, pp. 1003-1017, 2012.

\bibitem{}\label{korus2016}
P. Korus and J. Huang, ``Multi-scale fusion for improved localization of malicious tampering in digital images.'' \emph{IEEE Trans. Image Process.}, vol. 25, no. 3, pp. 1312-1326, 2016.

\bibitem{}\label{ferrara2012}
P. Ferrara, T. Bianchi, A. De Rosa and A. Piva, ``Image Forgery Localization via Fine-Grained Analysis of CFA Artifacts,'' \emph{IEEE Trans. Inf. Forensics and Security}, vol. 7, no. 5, pp. 1566-1577, 2012.

\bibitem{}\label{chierchia2014}
G. Chierchia, G. Poggi, C. Sansone and L. Verdoliva, ``A Bayesian MRF approach for PRNU-based image forgery detection,'' \emph{IEEE Trans. Inf. Forensics and Security}, vol. 9, no. 4, pp. 554-567, 2014.

\bibitem{}\label{korus2017}
P. Korus and J. Huang, ``Multi-Scale Analysis Strategies in PRNU-Based Tampering Localization, '' \emph{IEEE Trans. Inf. Forensics and Security}, vol. 12, no. 4, pp. 809-824, 2017.

\bibitem{}\label{bahrami2015}
K. Bahrami, A.C. Kot, L. Li and H. Li, ``Blurred image splicing localization by exposing blur type inconsistency,'' \emph{IEEE Trans. Inf. Forensics and Security}, vol. 10, no. 5, pp. 999-1009, 2015.

\bibitem{}\label{carvalho2016}
T. Carvalho, F. A. Faria, H. Pedrini, R. da S. Torres and A. Rocha, ``Illuminant-Based Transformed Spaces for Image Forensics,'' \emph{IEEE Trans. Inf. Forensics and Security}, vol. 11, no. 4, pp. 720-733, 2016.

\bibitem{}\label{zhang2009}
W. Zhang, X. Cao, Z. Feng, J. Zhang and P. Wang, ``Detecting photographic composites using two-view geometrical constraints,'' \emph{in Procs. IEEE Int. Conf. Multimedia Expo (ICME)}, pp. 1078-1081, 2009.

\bibitem{}\label{zhang2010}
W. Zhang, X. Cao, Y. Qu, Y. Hou, H. Zhao and C. Zhang, ``Detecting and extracting the photo composites using planar homography and graph-cut,'' \emph{IEEE Trans. Inf. Forensics and Security}, vol. 5, no. 3, pp. 544-555, 2010.

\bibitem{}\label{trung2014}
D.T. Trung, A. Beghdadi and M.-C. Larabi, ``Blind inpainting forgery detection,'' \emph{in Procs. IEEE Global Conf. Signal and Inf. Process. (GlobalSIP)}, pp.1019-1023, 2014.

\bibitem{}\label{Levin2004}
A. Levin, D. Lischinski and Y.Weiss, ``Colorization using optimization,'' \emph{ACM Trans. Graphics}, vol. 23, no. 3, pp. 689-694, 2004.

\bibitem{}\label{Pang2013}
J. Pang, O.C. Au, K. Tang and Y. Guo, ``Image colorization using sparse representation,'' \emph{in Proc. IEEE Int. Conf. Acoustics, Speech and Signal Process. (ICASSP)}, pp. 1578-1582, 2013.

\bibitem{}\label{Charpiat2008}
G. Charpiat, M. Hofmann and B. Scholkopf, ``Automatic image colorization via multimodal predictions,'' \emph{in Proc. European Conf. Comp. Vision (ECCV)}, pp. 126-139, 2008.

\bibitem{}\label{chen2016}
X. Chen, J. Li, D. Zou and Q. Zhao, ``Learn Sparse Dictionaries for Edit Propagation,'' \emph{IEEE Trans. Image Process.}, vol. 25, no. 4, pp. 1688-1698, 2016.

\bibitem{}\label{Cheng2015}
Z. Cheng, Q. Yang and B. Sheng, ``Deep colorization,'' \emph{in Proc. IEEE Int. Conf. Comp. Vision (ICCV)}, pp. 415-423, 2015.

\bibitem{}\label{lizuka2016tog}
S. Lizuka, E. Simo-Serra and H. Ishikawa,, ``Let there be color!: joint end-to-end learning of global and local image priors for automatic image colorization with simultaneous classification,'' \emph{ACM Trans. Graphics}, vol. 35, no. 4, pp. 110:1-110:11, 2016.

\bibitem{}\label{zhang2016eccv}
R. Zhang, P. Isola and A.A. Efros, ``Colorful image colorization,''  \emph{in Proc. European Conf. Comp. Vision (ECCV)}, pp. 649-666, 2016.

\bibitem{}\label{he2011dcp}
K. He, J. Sun and X. Tang, ``Single image haze removal using dark channel prior,'' \emph{IEEE Trans. Pattern Analysis and Machine Intelligence}, vol. 33, no. 12, pp. 2341-2353, 2011.


\bibitem{}\label{chang2011svm}
C.-C. Chang and C.-J. Lin, ``LIBSVM: a library for support vector machines,'' \emph{ACM Trans. Intelligent Systems and Technology}, vol. 2, no. 3, pp. 27:1-27:27, 2011.

\bibitem{}\label{vlfeat}
VLFeat: http://www.vlfeat.org/about.html

\bibitem{}\label{feifeili2009}
O. Russakovsky, J. Deng, H. Su, J. Krause, S. Satheesh, S. Ma, Z. Huang, A. Karpathy, A. Khosla, M. Bernstein, A.C. Berg and F.-F. Li, ``Imagenet large scale visual recognition challenge,'' \emph{Int. Journal of Computer Vision}, vol. 115, no. 3, pp. 211-252, 2015.

\bibitem{}\label{zisserman2007}
J. Philbin, O. Chum, M. Isard, J. Sivic and A. Zisserman, ``Object retrieval with large vocabularies and fast spatial matching,'' \emph{in Procs. IEEE Int. Conf. Comp. Vision and Pattern Recognition (CVPR)}, pp. 1-8, 2007.

\end{thebibliography}
\end{document}